\documentclass[a4paper,11pt]{article}
\usepackage{amsmath}
\usepackage{slashed}
\usepackage{amssymb}
\usepackage{epsfig}
\usepackage{graphicx}
\usepackage{multirow}
\usepackage{color}
\usepackage[normal]{subfigure}
\usepackage{rotating}
\usepackage{hyperref}
\usepackage[margin=0.9in]{geometry}
\usepackage{xcolor}
\usepackage{enumitem}
\usepackage{cite}

\def\eq#1{{Eq.~(\ref{#1})}}
\def\eqs#1#2{{Eqs.~(\ref{#1})--(\ref{#2})}}
\def\fig#1{{Fig.~\ref{#1}}}

\def\Table#1{{Table~\ref{#1}}}

\def\sect#1{{Sect.~\ref{#1}}}

\def\app#1{{Appendix~\ref{#1}}}

\def\abs#1{\left| #1\right|}

\def\Tr{\mbox{Tr}\,}
\def\det{\mbox{det}\,}

\begin{document}

\begin{flushright} 
TTP14-011, SFB/CPP-14-22
\end{flushright}

\vspace{1cm}
\begin{center}
{\huge\bf On the gauge dependence of the \\
\bigskip
Standard Model vacuum instability scale}\\
\bigskip\color{black}\vspace{0.6cm}
{
{\large  Luca Di Luzio}
{\large and Luminita Mihaila}
} \\[7mm]
{\it Institut f\"{u}r Theoretische Teilchenphysik,
Karlsruhe Institute of Technology (KIT), D-76128 Karlsruhe, Germany}\\[3mm]
\end{center}
\bigskip
\bigskip

\centerline{\large\bf Abstract}
\begin{quote}
\large
After reviewing the calculation of the Standard Model one-loop effective potential in a class of linear gauges, 
we discuss the physical observables entering the vacuum stability analysis. 
While the electroweak-vacuum-stability bound on the Higgs boson mass can be formally proven to be 
gauge independent, the field value at which the effective potential turns negative 
(the so-called instability scale) is a gauge dependent quantity. 
By varying the gauge-fixing scheme and the gauge-fixing parameters in their perturbative domain, 
we find an irreducible theoretical uncertainty of at least two orders 
of magnitude on the scale at which the Standard Model vacuum becomes unstable. 
\end{quote} 

\clearpage

\section{Introduction}

With the discovery of a Higgs-like boson at the LHC \cite{Aad:2012tfa,Chatrchyan:2012ufa}, the question of the 
Standard Model (SM) vacuum stability has received a renewed attention, 
with several high-precision analysis on the 
subject \cite{Holthausen:2011aa,EliasMiro:2011aa,Bezrukov:2012sa,Degrassi:2012ry,Alekhin:2012py,
Masina:2012tz,Buttazzo:2013uya,Antipin:2013sga,Branchina:2013jra} (see 
also \cite{Lindner:1988ww,Arnold:1989cb,Sher:1988mj,Sher:1993mf,Ford:1992mv,Altarelli:1994rb,Casas:1994qy,
Espinosa:1995se,Casas:1996aq,Isidori:2001bm,Isidori:2007vm,Ellis:2009tp}
for earlier works).
Absolute vacuum stability bounds  
are usually obtained by requiring that the electroweak vacuum is the absolute 
minimum of the effective potential, at least up to some cutoff scale, $\Lambda_{\rm{SM}}$, 
where the SM is not valid anymore 
and new physics is required in order to modify the shape of the effective 
potential.\footnote{Such a requirement can be relaxed if the tunnelling probability 
of the electroweak vacuum 
is small enough to comply with the age of the universe.}
It would be tempting (as it is often done) to identify the physical threshold, 
$\Lambda_{\rm{SM}}$, with the SM vacuum instability scale, $\Lambda$, 
which is operatively defined by the field value 
at which the effective potential becomes deeper than the electroweak minimum. 
However, due to the gauge dependence of the effective potential, 
$\Lambda$ suffers from an irreducible gauge ambiguity which makes its 
identification with $\Lambda_{\rm{SM}}$ problematic. 

The gauge dependence of the effective potential is known since long. 
Soon after the seminal work of Coleman and Weinberg \cite{Coleman:1973jx}, 
it was realized by Jackiw \cite{Jackiw:1974cv} that the effective potential is actually gauge dependent, 
thus raising the question of its physical significance. 
Since then, many authors have dealt with this subject 
\cite{Dolan:1974gu,Kang:1974yj,Fischler:1974ue,Frere:1974ia,Nielsen:1975fs,Fukuda:1975di,
Aitchison:1983ns,Johnston:1984sc,Thompson:1985hp,Kobes:1990dc,Ramaswamy:1995np,Metaxas:1995ab,DelCima:1999gg,Gambino:1999ai,Alexander:2008hd} 
and it is now a well-established practice to extract the physical content of the effective potential 
by means of the so-called Nielsen identities \cite{Nielsen:1975fs}. \\
In particular, the issue of the gauge dependence of the effective potential in the analysis of the SM vacuum stability 
was already pointed out at the end of the 90's 
by Loinaz and Willey \cite{Loinaz:1997td}, which challenged the possibility of setting gauge-independent 
lower bounds on the Higgs boson mass from vacuum stability constraints. 
More recently, the problematic identification between the cutoff scale of the SM and the 
instability scale $\Lambda$ was mentioned again in Ref.~\cite{Gonderinger:2012rd}. 

The aim of this paper is to clarify some issues related to the gauge dependence of the quantities entering the vacuum 
stability analysis. 
While the critical value of the Higgs boson mass, 
marking the transition between the stable and unstable phase of the SM, can be formally proven 
to be gauge independent, 
the SM instability scale is actually gauge dependent.  
This is explicitly shown by a direct calculation of the gauge dependent one-loop effective potential in the SM. \\
The SM effective potential is known in the Landau gauge at one \cite{Coleman:1973jx} 
and two loops \cite{Ford:1992pn,Martin:2001vx}  
since long. Recently, even the three-loop QCD and top-Yukawa corrections have been included \cite{Martin:2013gka}. 
On the other hand, calculations of the SM effective potential beyond the Landau gauge are less explored. 
Barring few exceptions, like for instance in Ref.~\cite{Patel:2011th} where a 
background-field-dependent gauge fixing with a single gauge-fixing parameter
was employed, the gauge dependence of the  SM effective potential is usually
not taken into consideration.

The paper is organized as follows: in \sect{SMoneloopEP} we provide a pedagogical derivation 
of the SM one-loop effective potential in the Fermi gauge (generalized Lorentz gauge) 
and consider its renormalization group (RG)
improvement. In \sect{physobsvacstab} we discuss the physical observables entering 
the vacuum stability analysis. In particular, by using the Nielsen identity \cite{Nielsen:1975fs}, 
we formally prove that the lower bound on the Higgs boson mass derived from the electroweak-vacuum-stability 
condition is gauge independent. On the other hand, the extrema of the effective potential and, in particular, 
the instability scale are in general gauge dependent. In \sect{gaugedepSMinst} we numerically quantify  
at the next-to-leading order (NLO) accuracy 
the gauge dependence of $\Lambda$ in the Fermi gauge by varying the gauge-fixing parameters 
in their perturbative domain and comment on the gauge-fixing scheme dependence of $\Lambda$. 
The interpretation and the physical implications of the gauge dependence 
of $\Lambda$ are discussed in \sect{conclusions}. The two-loop renormalization group equations (RGEs) of the 
SM parameters in the Fermi gauge are collected in \app{RGEapp}, while in \app{BCKGgaugefull} 
we report on the calculation of the SM one-loop effective potential in a background $R_\xi$ gauge 
with the most general set of gauge-fixing parameters. As a by-product we also obtain the SM one-loop effective 
potential in the standard $R_\xi$ gauge, whose expression might be useful for broken-phase calculations.

\section{The SM effective potential at one loop}
\label{SMoneloopEP}

In order to set the notation, let us split the classical Lagrangian density of
the electroweak sector of the SM in a gauge, Higgs and fermion part 
\begin{equation}
\label{Lclassical}
\mathcal{L}_{\rm{C}} = \mathcal{L}_{\rm{YM}} + \mathcal{L}_{\rm{H}} + \mathcal{L}_{\rm{F}} \, ,  
\end{equation} 
with 
\begin{align}
\label{LYM}
\mathcal{L}_{\rm{YM}} &=  
-\frac{1}{4} \left( \partial_\mu W^a_\nu - \partial_\nu W^a_\mu  + g \epsilon^{abc} W^b_\mu W^c_\nu \right)^2
-\frac{1}{4} \left( \partial_\mu B_\nu - \partial_\nu B_\mu  \right)^2 \, , \\
\label{LH}
\mathcal{L}_{\rm{H}} &= \left( D_\mu H \right)^\dag \left( D^\mu H \right) - V(H) \, , \\
\label{LF}
\mathcal{L}_{\rm{F}} &= \overline{Q}_L i \gamma_\mu D^\mu Q_L + \overline{t}_R i \gamma_\mu D^\mu t_R + 
\left(- y_t \overline{Q}_L (i\sigma^2) H^* t_R + \rm{h.c.}\right) + \ldots \, , 
\end{align}
where $W^a_\mu$ ($a=1,2,3$) and $B_\mu$ are the SU(2) and U(1) gauge fields, 
$H$ is the SM Higgs doublet with hypercharge $Y=1$ and $Q_L^T = (t_L, b_L)$ is the left-handed 
third generation quark doublet. 
Only the top quark is retained among the fermions and the QCD indices are suppressed in the quark sector.  
The covariant derivative is defined as 
\begin{equation}
\label{defcovder}
D_\mu = \partial_\mu - i g \frac{\sigma^a}{2} W^a_\mu + i g' \frac{Y}{2} B_\mu \, ,
\end{equation}
where $\sigma^a$ ($a=1,2,3$) are the usual Pauli matrices 
and with the term involving $g$ being absent for right-handed fermions. 
The Higgs potential is
\begin{equation}
V (H) = -m^2 H^\dag H + \lambda (H^\dag H)^2 \, . 
\end{equation}
The effective potential can be conveniently computed by means of the background field method of Jackiw \cite{Jackiw:1974cv}. 
After homogeneously shifting the scalar fields of the theory by a background (spacetime independent) field $\phi$, 
the one-loop effective potential is obtained by directly evaluating the path integral expression of the effective action in the Gaussian approximation. 
After some standard manipulations (see e.g.~also \cite{Delaunay:2007wb,Patel:2011th}), 
the one-loop effective potential 
\begin{equation}
\label{1loopEP}
V_{\rm{eff}}^{\rm{1-loop}} (\phi) =  V^{(0)}_{\rm{eff}} (\phi) + V^{(1)}_{\rm{eff}} (\phi) \, , 
\end{equation}
can be recast in terms of the well-known formulas \cite{Jackiw:1974cv}
\begin{align}
\label{treelevelEP}
V^{(0)}_{\rm{eff}} (\phi) &= V(\phi) \, , \\
\label{1loopEPclosed}
V^{(1)}_{\rm{eff}} (\phi) &= i \sum_{n \, 
= \, \text{SM fields}} \eta \int \frac{d^4 k}{(2\pi)^4} \log \det i \tilde{\mathcal{D}}^{-1}_n \{ \phi; k \} \, .
\end{align}
The matrix $i \tilde{\mathcal{D}}^{-1}_n \{ \phi; k \}$ denotes the $\phi$-dependent 
inverse propagators of the SM fields in momentum space, 
the determinant acts on all the internal indices  and $\eta = -1/2 \ (1)$ for bosons (fermions/ghosts) is the 
power of the functional determinant due to the Gaussian path integral. 

Gauge invariance allows us to perform the shift of the Higgs doublet in a specific direction 
of the $\rm{SU(2)} \otimes \rm{U(1)}$ space: 
\begin{equation}
\label{Hshift}
H(x) \rightarrow 
\frac{1}{\sqrt{2}}
\left( 
\begin{array}{c}
\chi^1(x) + i \chi^2(x) \\
\phi + h(x) + i \chi^3(x)
\end{array} 
\right)
\, ,  
\end{equation}
where $h$ denotes the Higgs field and $\chi^a$ ($a=1,2,3$) the Goldstone boson fields.
At tree level, the effective potential reads 
\begin{equation}
\label{V0}
V^{(0)}_{\rm{eff}} (\phi) = -\frac{m^2}{2} \phi^2 + \frac{\lambda}{4} \phi^4 \, ,
\end{equation} 
while in order to compute the quantum correction, $V^{(1)}_{\rm{eff}}$, one needs to work out the 
inverse propagators of the dynamical fields in the shifted SM Lagrangian.
For exemplification, we consider in the next section the computation of the one-loop
SM effective potential in the Fermi gauge. The calculation of the SM effective potential 
in a background-field-dependent $R_\xi$ gauge and in the standard $R_\xi$ gauge 
is instead presented in \app{BCKGgaugefull}.

\subsection{Fermi gauge}
\label{Fermigauge}

As long as we are interested in the high-energy behaviour of the the effective potential, 
we can directly work in the unbroken phase of the SM. 
Then, the most convenient way to fix the gauge 
is by means of the Fermi gauge (generalized Lorentz gauge): 
\begin{equation}
\label{gflagFermi}
\mathcal{L}^{\rm{Fermi}}_{\rm{g.f.}}  = -\frac{1}{2 \xi_W} \left( \partial^\mu W^a_\mu \right)^2 
-\frac{1}{2 \xi_B} \left( \partial^\mu B_\mu \right)^2 \, . 
\end{equation}
We are thus interested in the determination of the quadratic ($\phi$-dependent) part 
of the Lagrangian, $\mathcal{L}_{\rm{C}} + \mathcal{L}^{\rm{Fermi}}_{\rm{g.f.}}$, after the shift 
in \eq{Hshift}.\footnote{One can easily see that the bilinear ghost terms are $\phi$-independent. 
Hence, in the Fermi gauge the ghost contribution decouples from the one-loop effective potential.}
A straightforward calculation yields 
\begin{align}
\label{LYMquadFermi}
\mathcal{L}^{\rm{quad}}_{\rm{YM}} &= 
\tfrac{1}{2} W^a_\mu \left( \Box \, g^{\mu\nu} - \partial^\mu \partial^\nu \right) \delta^{ab} W^b_\nu 
+ \tfrac{1}{2} B_\mu \left( \Box \, g^{\mu\nu} - \partial^\mu \partial^\nu \right) B_\nu 
\, , \\
\label{LHquadFermi}
\mathcal{L}^{\rm{quad}}_{\rm{H}} &= 
\tfrac{1}{2} h \left( - \Box - \bar{m}_h^2 \right) h 
+ \tfrac{1}{2} \chi^a \left( - \Box - \bar{m}_\chi^2 \right) \delta^{ab} \chi^b  + \tfrac{1}{2} \bar{m}_W^2 W^a_\mu W^{a\mu}
+ \tfrac{1}{2} \bar{m}_B^2 B_\mu B^{\mu} 
\nonumber \\
& + \bar{m}_W \bar{m}_B W^3_\mu B^{\mu}  - \bar{m}_W \partial_\mu \chi^1 W^{2\mu} 
- \bar{m}_W \partial_\mu \chi^2 W^{1\mu} 
+ \bar{m}_W \partial_\mu \chi^3 W^{3\mu} 
+ \bar{m}_B \partial_\mu \chi^3 B^{\mu} \, , \\ 
\label{LFquadFermi}
\mathcal{L}^{\rm{quad}}_{\rm{F}} &= \overline{t} \left( i \slashed{\partial} - \bar{m}_t \right) t + \ldots \, , 
\end{align}
where $\Box \equiv \partial_\mu \partial^\mu$ and we defined the $\phi$-dependent masses
\begin{align}
\label{mhphiFermi}
\bar{m}_h^2 &= -m^2 + 3 \lambda \phi^2 \, , \\
\label{mchiphi}
\bar{m}_{\chi}^2 &= -m^2 + \lambda \phi^2  \, , \\
\label{defmwFermi}
\bar{m}_W &= \tfrac{1}{2} g \phi \, , \\
\label{defmbFermi}
\bar{m}_B &= \tfrac{1}{2} g' \phi \, , \\
\label{mtphiFermi}
\bar{m}_t &= \frac{y_t}{\sqrt{2}} \phi \, ,
\end{align}
while $\mathcal{L}^{\rm{Fermi}}_{\rm{g.f.}}$ is already quadratic in the gauge boson fields. 
The only technical complication in the Fermi gauge is the presence of a 
Goldstone--gauge boson mixing already at tree level (cf.~\eq{LHquadFermi}). 
The latter can be treated by defining an extended field vector 
\begin{equation}
\label{extX}
X^T = \left(V^T_\mu , \chi^T\right) 
\, ,
\end{equation}
where 
\begin{equation}
\label{defVchiFermi}
V^T_\mu = \left( W^1_\mu, W^2_\mu, W^3_\mu, B_\mu \right) \, , \qquad 
\chi^T = \left( \chi^1, \chi^2, \chi^3 \right) \, . 
\end{equation}
Then the quadratic part of the Goldstone--gauge sector can be rewritten
as
\begin{equation}
\label{quadgoldgauge}
\frac{1}{2} X^T \left( i \mathcal{D}_X^{-1} \right) X 
= \frac{1}{2} \left( V_\mu^T, \chi^T \right) 
\left(
\begin{array}{cc}
i \left( \mathcal{D}^{-1}_V \right)^\mu_{\nu} & \bar{m}^T_{\text{mix}} \, \partial^\mu  \\
- \bar{m}_{\text{mix}} \, \partial_\nu & i \mathcal{D}^{-1}_\chi
\end{array}
\right)
\left(
\begin{array}{c}
V^\nu \\
\chi
\end{array}
\right)
\, , 
\end{equation}
with 
\begin{equation}
\label{defmmix}
\bar{m}_{\text{mix}} = 
\left(
\begin{array}{cccc}
0 & - \bar{m}_W & 0 & 0 \\
- \bar{m}_W & 0 & 0 & 0 \\
0 & 0 & \bar{m}_W & \bar{m}_B
\end{array}
\right)
\, .  
\end{equation}
After Fourier transformation, $\partial_\mu \rightarrow i k_\mu$, the mixed inverse propagator matrix becomes 
\begin{equation}
\label{invpropX}
i \tilde{\mathcal{D}}_X^{-1} = 
\left(
\begin{array}{cc}
i ( \tilde{\mathcal{D}}^{-1}_V )^\mu_{\nu} & i k^\mu \bar{m}^T_{\text{mix}}  \\
- i k_\nu \bar{m}_{\text{mix}} & i \tilde{\mathcal{D}}^{-1}_\chi
\end{array}
\right) \, ,
\end{equation}
where $(\tilde{\mathcal{D}}^{-1}_V)^\mu_{\nu}$ is conveniently 
split into a transversal and a longitudinal part 
\begin{equation}
\label{invpropgaugeFermi}
( \tilde{\mathcal{D}}^{-1}_V )^\mu_\nu  =
i \tilde{\mathcal{D}}^{-1}_{T} \,(\Pi_{T})^\mu_\nu
+ i \tilde{\mathcal{D}}^{-1}_{L} \,(\Pi_{L})^\mu_\nu
\, , 
\end{equation}
with
\begin{equation}
\label{defPiTLFermi}
(\Pi_{T})^\mu_\nu = g^\mu_\nu - \frac{k^\mu k_\nu}{k^2} \, , 
\qquad
(\Pi_{L})^\mu_\nu = \frac{k^\mu k_\nu}{k^2} \, , 
\end{equation}
and  
\begin{align}
\label{invpropTFermi}
i \tilde{\mathcal{D}}^{-1}_{T} &= 
\left(
\begin{array}{cccc}
- k^2 + \bar{m}_W^2 & 0 & 0 & 0 \\
0 & - k^2 + \bar{m}_W^2 & 0 & 0 \\
0 & 0 & - k^2 + \bar{m}_W^2 & \bar{m}_W \bar{m}_B \\
0 & 0 & \bar{m}_W \bar{m}_B & - k^2 + \bar{m}_B^2
\end{array}
\right) \, , \\
\label{invpropLFermi}
i \tilde{\mathcal{D}}^{-1}_{L} &= 
\left(
\begin{array}{cccc}
- \xi_W^{-1} k^2 + \bar{m}_W^2  & 0 & 0 & 0 \\
0 & - \xi_W^{-1} k^2 + \bar{m}_W^2 & 0 & 0 \\
0 & 0 & - \xi_W^{-1} k^2 + \bar{m}_W^2  & \bar{m}_W \bar{m}_B \\
0 & 0 & \bar{m}_W \bar{m}_B & -\xi_B^{-1} k^2 + \bar{m}_B^2 
\end{array}
\right) \, . 
\end{align}
The Goldstone boson inverse propagator reads 
\begin{equation}
\label{invpropchiFermi}
i \tilde{\mathcal{D}}^{-1}_{\chi} = 
\left(
\begin{array}{ccc}
k^2 - \bar{m}_\chi^2 & 0 & 0 \\
0 & k^2 - \bar{m}_\chi^2  & 0 \\
0 & 0 &k^2 - \bar{m}_\chi^2  
\end{array}
\right) \, ,
\end{equation}
while those of the Higgs and top quark fields are
\begin{align}
\label{invprophFermi}
i \tilde{\mathcal{D}}^{-1}_{h} &= 
k^2 - \bar{m}_h^2 \, , \\
\label{invproptFermi}
i \tilde{\mathcal{D}}^{-1}_{t} &= \slashed{k} - \bar{m}_t \, . 
\end{align}
The next step (see \eq{1loopEPclosed}) is the evaluation of $\log\det i \tilde{\mathcal{D}}_n^{-1}$, for $n = X, h ,t$. 
Only the former and the latter present some non-trivial steps. 
Let us start by expressing the determinant of the block matrix in \eq{invpropX} as  
\begin{align}
\label{evaluationdetX}
\det i \tilde{\mathcal{D}}_X^{-1} &= \det i \tilde{\mathcal{D}}_\chi^{-1} \det 
\left( 
i ( \tilde{\mathcal{D}}^{-1}_V )^\mu_{\nu} 
- k^\mu k_\nu \bar{m}^T_{\text{mix}} \left( i \tilde{\mathcal{D}}^{-1}_{\chi} \right)^{-1} \bar{m}_{\text{mix}}
\right) \nonumber \, , \\
&= \det i \tilde{\mathcal{D}}_\chi^{-1} \det 
\left( 
i \tilde{\mathcal{D}}^{-1}_{T} (\Pi_{T})^\mu_\nu  
+ \left( i \tilde{\mathcal{D}}^{-1}_{L}  - k^2 \bar{m}^T_{\text{mix}} \left( i \tilde{\mathcal{D}}^{-1}_{\chi} \right)^{-1} \bar{m}_{\text{mix}} \right) (\Pi_{L})^\mu_\nu
\right) \, , 
\end{align}
where in the last step we used \eq{invpropgaugeFermi},
and perform a Lorentz transformation in $d$ spacetime dimensions,\footnote{We
  already anticipate the fact that we are going to regulate the divergent
  integrals in  
dimensional regularization.} $k_\mu \rightarrow (k_0, 0, 0, 0, \ldots)$, such
that $(\Pi_{L})^\mu_\nu \rightarrow (1,0,0,0,\ldots)$  
and $(\Pi_{T})^\mu_\nu \rightarrow (0,1,1,1,\ldots)$. 
Using the Loretz invariance of the determinant, we obtain
\begin{equation}
\label{evaluationlogdetX}
\log\det i \tilde{\mathcal{D}}_X^{-1} =  
(d-1) \log \det \tilde{\mathcal{D}}^{-1}_{T}  
+ \log \det i \tilde{\mathcal{D}}_\chi^{-1} 
\det \left( i \tilde{\mathcal{D}}^{-1}_{L}  - k^2 \bar{m}^T_{\text{mix}} \left( i \tilde{\mathcal{D}}^{-1}_{\chi} \right)^{-1} \bar{m}_{\text{mix}} \right) \, . 
\end{equation} 
The explicit evaluation of the two summands in the right-hand side of \eq{evaluationlogdetX} yields
\begin{equation}
\label{logdetTFermi}
\log \det i \tilde{\mathcal{D}}^{-1}_T = 
2 \log \left( -k^2 + \bar{m}_W^2 \right) 
+ \log \left( -k^2 + \bar{m}_{Z}^2 \right)
+ \ldots \, , \\ 
\end{equation}
and 
\begin{align}
\label{rhsexpl}
& \log \det i \tilde{\mathcal{D}}_\chi^{-1} 
\det \left( i \tilde{\mathcal{D}}^{-1}_{L}  
- k^2 \bar{m}^T_{\text{mix}} \left( i \tilde{\mathcal{D}}^{-1}_{\chi} \right)^{-1} \bar{m}_{\text{mix}} \right) \nonumber \\ 
& =
2 \log \left( k^4 -k^2 \bar{m}_\chi^2 + \bar{m}_\chi^2 \xi_W \bar{m}_W^2 \right) 
+ \log \left( k^4 -k^2 \bar{m}_\chi^2 + \bar{m}_\chi^2 (\xi_W \bar{m}_W^2 + \xi_B \bar{m}_B^2) \right) + \ldots \nonumber \\
& = 2 \log \left( k^2 - \bar{m}_{A^{+}}^2 \right)
+ 2 \log \left( k^2 - \bar{m}_{A^{-}}^2 \right) + \log \left( k^2 - \bar{m}_{B^{+}}^2 \right) + \log \left( k^2 - \bar{m}_{B^{-}}^2 \right) + \ldots \, ,
\end{align}
where the ellipses stand for $\phi$-independent terms and we defined the $\phi$-dependent masses
\begin{align} 
\label{defmassZFermi}
\bar{m}_{Z}^2 &= \bar{m}_W^2 + \bar{m}_B^2 \, , \\
\label{defmassApm}
\bar{m}_{A^{\pm}}^2 &= \frac{1}{2} \bar{m}_\chi \left(  \bar{m}_\chi \pm \sqrt{ \bar{m}_\chi^2 - 4 \xi_W \bar{m}_W^2} \right) \, , \\ 
\label{defmassBpm}
\bar{m}_{B^{\pm}}^2 &= \frac{1}{2} \bar{m}_\chi \left(  \bar{m}_\chi \pm \sqrt{ \bar{m}_\chi^2 
- 4 (\xi_W \bar{m}_W^2 + \xi_B \bar{m}_B^2) } \right) \, . 
\end{align}
For the evaluation of the fermionic determinant of \eq{invproptFermi} we employ 
a naive treatment of $\gamma_5$ in dimensional 
regularization (i.e.~$\{ \gamma_5, \gamma_\mu \} = 0$ in $d$ dimensions) and make the standard choice 
$\Tr \mathbf{1}_\text{Dirac} = 4$ in $d$ dimensions.\footnote{A different choice, e.g.~$\Tr \mathbf{1}_\text{Dirac} = 2^{d/2}$, 
would just lead to a different renormalization scheme \cite{Collins:1984xc}.} 
Explicitly, one has  
\begin{align}
\log \det \left( \slashed{k} - \bar{m}_t \right)
&= \Tr \log \left( \slashed{k} - \bar{m}_t \right)
= \Tr \log \gamma^5 \left( \slashed{k} - \bar{m}_t \right) \gamma^5
= \Tr \log \left( - \slashed{k} - \bar{m}_t \right) \nonumber \\
&= \frac{1}{2} \left[ \Tr \log \left( \slashed{k} - \bar{m}_t \right) + \Tr \log \left( - \slashed{k} - \bar{m}_t \right) \right]
= \frac{1}{2}  \Tr \log \left( - k^2 + \bar{m}_t^2 \right) \nonumber \\
&= \frac{1}{2} 4 \times 3 \log \left( -k^2 + \bar{m}_t^2 \right) \, ,
\end{align}
where the extra factors in the last step are due to the trace in the Dirac and color space. 

Including all the relevant degrees of freedom 
and working in dimensional regularization with $d = 4 - 2 \epsilon$, 
the one-loop contribution to the effective potential (cf.~again \eq{1loopEPclosed}) 
can be adjusted in the following way:
\begin{align}
\label{EP1loopSMdrexplFermi}
V^{(1)}_{\rm{eff}}(\phi)|^{\rm{Fermi}}  &= -\frac{i}{2} \mu^{2\epsilon} \int \frac{d^d k}{(2\pi)^d} 
\left[
- 12 \log \left( - k^2 + \bar{m}_t^2 \right) + (d-1) \left( 2 \log \left( -k^2 + \bar{m}_W^2 \right) 
\right. \right. \nonumber \\ 
& \left. + \log \left( -k^2 + \bar{m}_Z^2 \right) \right) + \log \left( k^2 - \bar{m}_h^2 \right) 
+ 2 \log \left( k^2 - \bar{m}_{A^{+}}^2 \right) + 2 \log \left( k^2 - \bar{m}_{A^{-}}^2 \right)  
\nonumber \\ 
& \left. + \log \left( k^2 - \bar{m}_{B^{+}}^2 \right) + \log \left( k^2 - \bar{m}_{B^{-}}^2 \right) + \ \text{$\phi$-independent} \right] \, .
\end{align} 
The integrals are easily evaluated after Wick rotation, yielding 
\begin{equation}
\label{integralFermi}
-\frac{i}{2} \mu^{2\epsilon} \int \frac{d^d k}{(2\pi)^d} \log (-k^2 + m^2) = 
\frac{1}{4} \frac{m^4}{(4\pi)^2} \left( \log \frac{m^2}{\mu^2} -\frac{3}{2} -\Delta_\epsilon \right)
\, , 
\end{equation}
where we introduced the modified minimal subtraction ($\overline{\rm{MS}}$) term \cite{Bardeen:1978yd}
\begin{equation}
\label{defDeltaepsFermi}
\Delta_\epsilon = \frac{1}{\epsilon} - \gamma_E + \log 4 \pi \, .
\end{equation}  
After the $\epsilon$-expansion
the one-loop contribution to the effective potential is given by
\begin{align}
\label{1loopEPbareFermi} 
& V^{(1)}_{\rm{eff}}|_{\rm{bare}}^{\rm{Fermi}} = \frac{1}{4 (4 \pi)^2} \left[ 
-12 \bar{m}_t^4 \left( \log\frac{\bar{m}_t^2}{\mu^2} - \frac{3}{2} - \Delta_\epsilon \right)
+6 \bar{m}_W^4 \left( \log\frac{\bar{m}_W^2}{\mu^2} - \frac{5}{6} - \Delta_\epsilon \right) \right. \\
& \left.+3 \bar{m}_Z^4 \left( \log\frac{\bar{m}_Z^2}{\mu^2} - \frac{5}{6} - \Delta_\epsilon \right) 
+\bar{m}_h^4 \left( \log\frac{\bar{m}_h^2}{\mu^2} - \frac{3}{2} - \Delta_\epsilon \right)
+2 \bar{m}_{A^+}^4 \left( \log\frac{\bar{m}_{A^+}^2}{\mu^2} - \frac{3}{2} - \Delta_\epsilon \right)  \right. \nonumber \\
& \left. +2 \bar{m}_{A^-}^4 \left( \log\frac{\bar{m}_{A^-}^2}{\mu^2} - \frac{3}{2} - \Delta_\epsilon \right) 
+ \bar{m}_{B^+}^4 \left( \log\frac{\bar{m}_{B^+}^2}{\mu^2} - \frac{3}{2} - \Delta_\epsilon \right)
+ \bar{m}_{B^-}^4 \left( \log\frac{\bar{m}_{B^-}^2}{\mu^2} - \frac{3}{2} - \Delta_\epsilon \right)
\right] \, . \nonumber 
\end{align}
In particular, in terms of the SM couplings the divergent part of \eq{1loopEPbareFermi} reads 
\begin{align}
\label{1loopdivFermi} 
V^{(1)}_{\rm{eff}}|_{\rm{bare-pole}}^{\rm{Fermi}} &= \frac{\Delta_\epsilon}{(4\pi)^2} 
\left[
-m^4
+ \left(3 \lambda - \frac{1}{8} \xi _B g'^2 - \frac{3}{8} \xi _W g^2   \right) m^2 \phi^2 \right. \nonumber \\
& \left. +
   \left(
   -\frac{3}{64} g'^4
   -\frac{3}{32} g'^2 g^2
   -\frac{9}{64} g^4
   +\frac{3}{4} y_t^4
   -3 \lambda ^2
   +\frac{1}{8} \xi _B g'^2 \lambda 
   +\frac{3}{8}  \xi _W g^2  \lambda 
   \right) \phi^4
\right] \, .
\end{align}
While the $m^4$-dependent pole in \eq{1loopdivFermi} can be always subtracted by a constant shift in the 
effective potential,\footnote{A constant shift in the effective potential does not affect the equations of 
motion, 
as long as gravity is ignored.} 
the remaining divergences are canceled by the multiplicative renormalization of the bare field and couplings 
appearing in $V^{(0)}_{\rm{eff}}$ (cf.~\eq{V0}): 
\begin{equation}
\label{renV0parFermi}
\phi_0 = Z_{\phi}^{1/2}|^{\rm{Fermi}} \phi \, , \qquad
m^2_0 = Z_{m^2} m^2 \, , \qquad 
\lambda_0 = Z_{\lambda} \lambda \, ,
\end{equation}
where the renormalization constants can be conveniently computed in the unbroken phase of the SM.  
Their expressions at one loop  in the $\overline{\rm{MS}}$ scheme read (see e.g.~\cite{Chetyrkin:2012rz,Mihaila:2012pz}):
\begin{align}
\label{ZphiFermi} 
Z_{\phi}^{1/2}|^{\rm{Fermi}}  &= 1 + 
\frac{\Delta_\epsilon}{(4\pi)^2} 
\left( 
\frac{3}{8} g'^2 + \frac{9}{8} g^2 -\frac{3}{2} y_t^2 -\frac{1}{8} \xi _B g'^2-\frac{3}{8} \xi _W g^2   
\right) \, , \\
\label{Zm2Fermi} 
Z_{m^2} &= 1+ \frac{\Delta_\epsilon}{(4\pi)^2} 
\left( 
-\frac{3}{4} g'^2 -\frac{9}{4} g^2 +3 y_t^2 +6 \lambda
\right) \, , \\
\label{ZlamFermi} 
Z_{\lambda} &= 1+ \frac{\Delta_\epsilon}{(4\pi)^2} 
\left(
 -\frac{3}{2} g'^2 -\frac{9}{2} g^2 +6 y_t^2 +12 \lambda
+\frac{3 }{16} \frac{g'^4}{\lambda }  
+\frac{3}{8} \frac{g'^2 g^2}{\lambda }
+\frac{9}{16}\frac{g^4}{\lambda }
-3 \frac{y_t^4}{\lambda }
\right) \, .
\end{align}
It is a simple exercise to check that the renormalization of the tree-level potential, 
via the renormalization constants in \eqs{ZphiFermi}{ZlamFermi}, cancels the
$\phi$-dependent poles in \eq{1loopdivFermi}. Let us point out that in the
Fermi gauge the field $\phi$
gets only multiplicatively renormalized by the wavefunction
of the Higgs field. This feature is due to the invariance 
of the complete SM Lagrangian (including the gauge-fixing term in \eq{gflagFermi}) 
under the transformation $h \rightarrow h + a$ and $\phi \rightarrow \phi - a$,  
as shown in \cite{Pilaftsis:1997fe,Binosi:2005yk}. 
As we will see in \app{BCKGgaugefull}, this property does not hold anymore in the background $R_\xi$ gauge. 

Hence, after the renormalization procedure, 
the one-loop contribution to the effective potential 
in the $\overline{\rm{MS}}$ scheme reads  
\begin{align}
\label{1loopEPFermi} 
V^{(1)}_{\rm{eff}}|^{\rm{Fermi}}  &= \frac{1}{4 (4 \pi)^2} \left[ 
-12 \bar{m}_t^4 \left( \log\frac{\bar{m}_t^2}{\mu^2} - \frac{3}{2} \right)
+6 \bar{m}_W^4 \left( \log\frac{\bar{m}_W^2}{\mu^2} - \frac{5}{6}  \right) \right. \nonumber \\
& \left.+3 \bar{m}_Z^4 \left( \log\frac{\bar{m}_Z^2}{\mu^2} - \frac{5}{6}  \right) 
+\bar{m}_h^4 \left( \log\frac{\bar{m}_h^2}{\mu^2} - \frac{3}{2}  \right)
+2 \bar{m}_{A^+}^4 \left( \log\frac{\bar{m}_{A^+}^2}{\mu^2} - \frac{3}{2} \right)  \right. \nonumber \\
& \left. +2 \bar{m}_{A^-}^4 \left( \log\frac{\bar{m}_{A^-}^2}{\mu^2} - \frac{3}{2}  \right) 
+ \bar{m}_{B^+}^4 \left( \log\frac{\bar{m}_{B^+}^2}{\mu^2} - \frac{3}{2}  \right)
+ \bar{m}_{B^-}^4 \left( \log\frac{\bar{m}_{B^-}^2}{\mu^2} - \frac{3}{2}  \right)
\right] \, , 
\end{align}
where the definitions of the $\phi$-dependent mass terms are given in 
\eqs{mhphiFermi}{mtphiFermi} and \eqs{defmassZFermi}{defmassBpm}.
In particular, for $\xi_W = \xi_B = 0$ one has $\bar{m}_{A^+} = \bar{m}_{B^+}
= \bar{m}_{\chi}$ and $\bar{m}_{A^-} = \bar{m}_{B^-} = 0$,  
so that \eq{1loopEPFermi} reproduces the standard one-loop result in the Landau gauge \cite{Coleman:1973jx}.  

Let us stress that the gauge dependence of $V^{(1)}_{\rm{eff}}$ cannot be removed 
by a suitable choice of the renormalization scheme, 
as it can be verified by adding finite terms in \eqs{ZphiFermi}{ZlamFermi}. Notice, however, 
that on the tree-level minimum, 
$m^2 = \lambda \phi^2$ (hence
$\bar{m}_{\chi} = 0$ and $\bar{m}_{A^\pm} = \bar{m}_{B^\pm} = 0$),  
the gauge dependence drops from $V^{(1)}_{\rm{eff}}|^{\rm{Fermi}} $. We
will discuss this aspect in more detail in \sect{physobsvacstab}.

\subsection{Renormalization group improvement}
\label{rengroup}

In applications where the behavior of $V_{\rm{eff}}(\phi)$ at large $\phi$ is needed, 
like for the vacuum stability analysis, one has to deal with potentially large logarithms 
of the type $\log (\phi/\mu)$ which may spoil the applicability range of perturbation theory. 
The standard way to resum such logarithms is by means of the RGEs. 
Since $V_{\rm{eff}}$ is independent of the renormalization scale $\mu$ for fixed values of the bare parameters, 
one obtains the RGE
\begin{equation}
\label{RGE}
\left( \mu \frac{\partial}{\partial \mu} + \beta_i \frac{\partial}{\partial \lambda_i} 
- \gamma \phi \frac{\partial}{\partial \phi} \right) V_{\rm{eff}} = 0 \, ,
\end{equation}
where the beta functions 
\begin{equation}
\label{defbf}
\beta_i = \mu \frac{d \lambda_i}{d \mu} \, ,
\end{equation}
correspond to each of the SM coupling $\lambda_i$ (including the gauge-fixing parameters) 
and the anomalous dimension of the background field is defined by 
\begin{equation}
\label{defanomdim}
\gamma = 
- \frac{\mu}{\phi} \frac{d \phi}{d \mu} \, .
\end{equation}
The formal solution of the RGE in \eq{RGE} can be obtained by applying the method of the characteristics \cite{Ford:1992mv}: 
\begin{equation}
\label{solRGE}
V_{\rm{eff}} (\mu, \lambda_i, \phi ) = V_{\rm{eff}}(\mu (t), \lambda_i (t), \phi (t) ) \, ,
\end{equation}
where
\begin{align}
\label{mut}
\mu(t) &= \mu e^t \, , \\
\label{phit}
\phi(t) &=  e^{\Gamma(t)} \phi \, , 
\end{align}
with
\begin{equation}
\label{Gammat}
\Gamma(t) = - \int_0^t \gamma(\lambda(t')) \, dt' \, ,
\end{equation}
and $\lambda_i (t)$ are the SM running couplings, determined by the equation 
\begin{equation}
\label{betat}
\frac{d \lambda_i (t)}{dt} = \beta_i (\lambda_i(t)) \, ,
\end{equation}
and subject to the boundary condition $\lambda_i (0) = \lambda_i$. 

The usefulness of the RG is that $t$ can be chosen in such a way that 
the convergence of perturbation theory is improved. 
For instance, a standard choice in vacuum stability analyses is $\mu (t) = \phi$ (see e.g.~Ref.~\cite{Degrassi:2012ry}). 
Without sticking, for the time being, to any specific choice of scale, 
the RG improved effective potential can be rewritten as
\begin{equation}
\label{1loopEPimproved}
V_{\rm{eff}} (\phi,t) = 
\Omega_{\rm{eff}}(\phi,t) -\frac{m_{\rm{eff}}^2(\phi,t)}{2} \phi^2 + \frac{\lambda_{\rm{eff}}(\phi,t)}{4} \phi^4 \, , 
\end{equation}
where the functional form of the effective couplings in \eq{1loopEPimproved} 
depends on the chosen gauge. 
In particular, in the limit $\phi \gg m$ the effective potential takes the universal form 
\begin{equation}
\label{1loopEPimprovedapprox}
V_{\rm{eff}} (\phi,t) \approx \frac{\lambda_{\rm{eff}}(\phi,t)}{4} \phi^4 \, , 
\end{equation}
with
\begin{equation}
\label{lameffapprox}
\lambda_{\rm{eff}}(\phi,t) \approx e^{4 \Gamma(t)} \left[ \lambda(t) 
+ \frac{1}{(4\pi)^2} \sum_p N_p \kappa_p^2 (t) 
\left( \log \frac{\kappa_p(t) e^{2 \Gamma(t)} \phi^2}{\mu(t)^2} - C_p \right) \right] \, , 
\end{equation}
since $\phi$ is the only massive parameter.
The coefficients $N_p$, $C_p$ and $\kappa_p$ appearing in \eq{lameffapprox} 
are explicitly listed in \Table{tab:pvaluesFermi} for the Fermi gauge 
and in \Table{tab:pvaluesBCKG} of \app{BCKGgaugefull} for the background $R_\xi$ gauge.
\begin{table*}[h]
  \begin{center}  
      \begin{tabular}{|c|cccccc|}
      \hline
        $p$ & $t$ & $W$ & $Z$ & $h$ & $A^{\pm}$ & $B^{\pm}$ \\ 
        \hline
        $N_p$ & $-12$ & $6$ & $3$ & $1$ & $2$ & $1$ \\ 
        $C_p$ & $\frac{3}{2}$ & $\frac{5}{6}$ & $\frac{5}{6}$ & $\frac{3}{2}$ & $\frac{3}{2}$ & $\frac{3}{2}$ \\ 
        $\kappa_p$ & $\frac{y_t^2}{2}$ & $\frac{g^2}{4}$ & $\frac{g^2+g'^2}{4}$ & $3 \lambda$ 
        & $\frac{1}{2} \left( \lambda \pm \sqrt{\lambda^2 - \lambda \xi_W g^2} \right)$ 
        & $\frac{1}{2} \left( \lambda \pm \sqrt{\lambda^2 - \lambda (\xi_W g^2 + \xi_B g'^2)} \right)$  \\ 
        \hline
        \end{tabular}
    \caption{\label{tab:pvaluesFermi} The $p$-coefficients entering the expression of $\lambda_{\rm{eff}}$ 
    in \eq{lameffapprox} for the Fermi gauge.
      }
  \end{center}
\end{table*}

Let us finally note that the gauge dependence of the RG improved effective potential is twofold. 
The gauge fixing parameters appear both in the couplings $\kappa_p$ (cf.~\Table{tab:pvaluesFermi}), 
and in the anomalous dimension of $\phi$ (cf.~\eq{RGE2lphi} in \app{RGEapp}) and hence in 
its integral $\Gamma$. 

\section{Physical observables in the vacuum stability analysis}
\label{physobsvacstab}

\begin{figure}[h]
\centering
\includegraphics[angle=0,width=12cm]{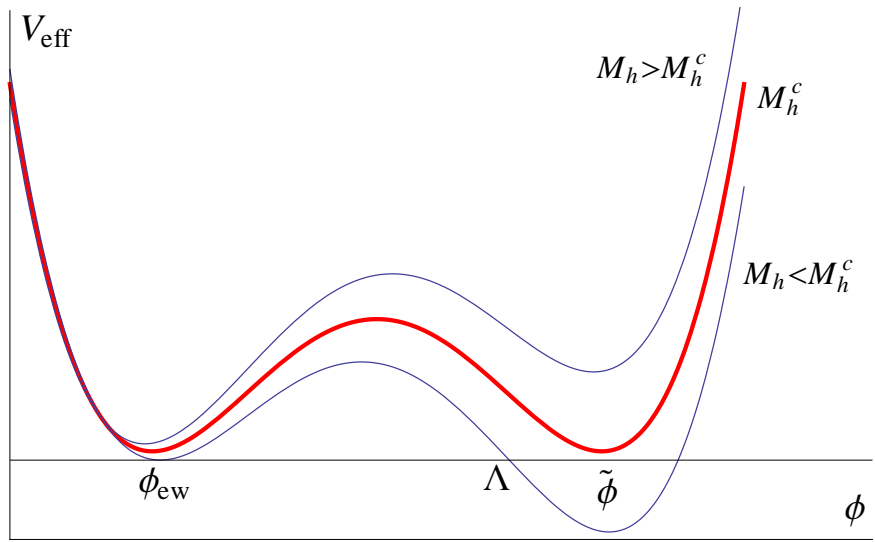}
\caption{\label{Mchplot} 
Schematic representation of the SM effective potential for different values of the Higgs boson mass. 
For $M_h < M_h^c$, the electroweak vacuum is unstable.}
\end{figure}

The present Section is devoted to a general discussion on the gauge dependence/independence 
of the quantities entering the 
vacuum stability analysis.  
To fix the ideas, let us assume that all the parameters of the SM are exactly determined, 
but the Higgs boson mass. 
After choosing the renormalization scale $t$, the RG improved effective potential, 
$V_{\rm{eff}} (\phi, M_h; \xi)$, is a function of $\phi$, the Higgs pole mass $M_h$, and the gauge fixing parameters, which are collectively denoted by $\xi$. 
One can think of $M_h$ as an order parameter, whose variation modifies the shape of the effective potential, as for instance sketched in \fig{Mchplot}.   

The absolute stability bound on the Higgs boson mass can be obtained by defining a ``critical'' mass, $M_h^c$, 
for which the value of the effective potential at the electroweak minimum, $\phi_{\rm{ew}}$, 
and at a second minimum, $\tilde{\phi} > \phi_{\rm{ew}}$, are the same. Analytically, 
this translates into the three conditions:
\begin{align}
\label{absvacstab1}
& V_{\rm{eff}} (\phi_{\rm{ew}}, M_h^c; \xi) - V_{\rm{eff}} (\tilde{\phi}, M_h^c; \xi) = 0 \, , \\ 
\label{absvacstab2}
& \left. \frac{\partial V_{\rm{eff}}}{\partial \phi} \right|_{\phi_{\rm{ew}}, M_h^c} = \left. \frac{\partial V_{\rm{eff}}}{\partial \phi} \right|_{\tilde{\phi}, M_h^c} = 0 \, . 
\end{align}
In the $\phi \gg \phi_{\rm{ew}}$ limit, the RG improved SM effective potential is well approximated by  
\begin{equation}
\label{Veffapprox}
V_{\rm{eff}} (\phi) = 
\left( \frac{\Omega_{\rm{eff}}(\phi)}{\phi^4} 
- \frac{1}{2} \frac{m_{\rm{eff}}^2 (\phi)}{\phi^2} 
+ \frac{1}{4} \lambda_{\rm{eff}}(\phi) \right)  \phi^4 \approx \frac{1}{4} \lambda_{\rm{eff}}(\phi) \phi^4 \, .
\end{equation} 
Indeed, at the leading order in the $m^2/\phi^2$ expansion, 
where $m^2 \sim \phi_{\rm{ew}}^2$ is the electroweak parameter of the Higgs potential, 
the effective couplings $\Omega_{\rm{eff}}$ and $m^2_{\rm{eff}}$ turn out to be 
proportional to $m^4$ and $m^2$ respectively.\footnote{
Moreover, since the beta function of $m$ is proportional to $m$ itself, 
the value of $m$ does not change much even after a scale running of many orders of magnitude.}
Hence, the absolute stability condition in \eqs{absvacstab1}{absvacstab2}
can be equivalently rewritten in the following way \cite{Bezrukov:2012sa}: 
\begin{align}
\label{absvacstab1mod}
\lambda_{\rm{eff}} (\tilde{\phi}, M_h^c; \xi) = 0 \, , \\
\label{absvacstab2mod}
\left. \frac{\partial \lambda_{\rm{eff}}}{\partial \phi} \right|_{\tilde{\phi}, M_h^c} = 0 \, , 
\end{align}
up to $\phi_{\rm{ew}}^2 / \tilde{\phi}^2 \ll 1$ corrections. 

On the other hand, due to the explicit presence of $\xi$ in the vacuum
stability condition, it is not obvious a priori which are  
the physical (gauge-independent) observables entering the vacuum stability analysis. 
The basic tool, in order to capture the gauge-invariant content of the effective potential is given by the 
Nielsen identity \cite{Nielsen:1975fs}
\begin{equation}
\label{NielsenIdbis}
\frac{\partial}{\partial \xi} V_{\text{eff}}(\phi,\xi) = - C(\phi, \xi)
\frac{\partial}{\partial \phi} V_{\text{eff}}(\phi,\xi)  \, ,  
\end{equation}
where $C(\phi, \xi)$ is a correlator involving the ghost fields and the
gauge-fixing functional, whose explicit expression  
will not be needed for our argument. \eq{NielsenIdbis} is valid for the class
of linear gauges and can be derived  
from the BRST non-invariance of a composite operator involving the ghost field and the gauge fixing functional 
(see e.g.~\cite{Metaxas:1995ab} for a concise derivation). 

The identity in \eq{NielsenIdbis} carries the following interpretation: the
effective potential is gauge independent where it is  
stationary and hence spontaneous symmetry breaking is a gauge-invariant statement. 
In the rest of this section we will use the Nielsen identity, in combination
with the vacuum stability condition in \eqs{absvacstab1}{absvacstab2},   
in order to formally prove that the critical Higgs boson mass, $M_h^c$, is a gauge-independent quantity, 
while the position of the extrema of the effective potential (e.g.~$\tilde{\phi}$) or the point where 
$V_{\rm{eff}}$ takes a special value (for instance zero) are essentially gauge dependent.

Our arguments are similar to those presented in Ref.~\cite{Patel:2011th}, 
about the gauge independence of the critical temperature of a first order phase transition 
in the context of the finite temperature effective potential.

\subsection{Gauge independence of the critical Higgs boson mass}
\label{gaugeindepMHc}

Let us assume that simultaneously inverting \eqs{absvacstab1}{absvacstab2} 
would yield gauge dependent field values and critical Higgs boson mass: 
$ \phi_{\rm{ew}} = \phi_{\rm{ew}} (\xi)$, $\tilde{\phi} = \tilde{\phi} (\xi)$ and $M_h^c = M_h^c (\xi)$.
The total differential of \eq{absvacstab1} with respect to $\xi$ then reads 
\begin{multline}
\label{totaldeiffxiA}
\left. \frac{\partial V_{\rm{eff}}}{\partial \phi} \right|_{\phi_{\rm{ew}}, M_h^c} \frac{\partial \phi_{\rm{ew}}}{\partial \xi}
+ \left. \frac{\partial V_{\rm{eff}}}{\partial M_h} \right|_{\phi_{\rm{ew}}, M_h^c} \frac{\partial M_h^c}{\partial \xi} 
+ \left. \frac{\partial V_{\rm{eff}}}{\partial \xi} \right|_{\phi_{\rm{ew}}, M_h^c} = \\
\left. \frac{\partial V_{\rm{eff}}}{\partial \phi} \right|_{\tilde{\phi}, M_h^c} \frac{\partial \tilde{\phi}}{\partial \xi}
+ \left. \frac{\partial V_{\rm{eff}}}{\partial M_h} \right|_{\tilde{\phi}, M_h^c} \frac{\partial M_h^c}{\partial \xi} 
+ \left. \frac{\partial V_{\rm{eff}}}{\partial \xi} \right|_{\tilde{\phi}, M_h^c} \, .
\end{multline}
The first term in both the left-hand side (lhs) and the right-hand side (rhs) 
of \eq{totaldeiffxiA} vanishes because of the stationary conditions in \eq{absvacstab2}. 
The third term in both the lhs and the rhs of \eq{totaldeiffxiA} vanishes for the same reason, after using the Nielsen identity.
Hence, we are left with 
\begin{equation}
\label{leftwith1}
\left( \left. \frac{\partial V_{\rm{eff}}}{\partial M_h} \right|_{\phi_{\rm{ew}}, M_h^c} 
- \left. \frac{\partial V_{\rm{eff}}}{\partial M_h} \right|_{\tilde{\phi}, M_h^c} \right)  \frac{\partial M_h^c}{\partial \xi} = 0 \, .
\end{equation}
Since the expression in the bracket of \eq{leftwith1} is in general different from zero, one concludes that 
\begin{equation}
\label{leftwith}
\frac{\partial M_h^c}{\partial \xi} = 0 \, ,
\end{equation}
namely, the critical Higgs boson mass is gauge independent. 
Let us notice, however, that the statement above formally holds at all orders in perturbation theory. 

\subsection{Gauge dependence of the extrema of the effective potential}

Let us consider now the total differential with respect to $\xi$ of the second expression in \eq{absvacstab2}
\begin{equation}
\label{totaldeiffxiB}
\left. \frac{\partial^2 V_{\rm{eff}}}{\partial \phi^2} \right|_{\tilde{\phi}, M_h^c} \frac{\partial \tilde{\phi}}{\partial \xi}
+ \left. \frac{\partial^2 V_{\rm{eff}}}{\partial M_h \ \partial \phi} \right|_{\tilde{\phi}, M_h^c} \frac{\partial M_h^c}{\partial \xi} 
+ \left. \frac{\partial^2 V_{\rm{eff}}}{\partial \xi \ \partial \phi} \right|_{\tilde{\phi}, M_h^c} = 0 \, .
\end{equation}
The second term is zero due to \eq{leftwith}. By differentiating the Nielsen identity with respect to $\phi$, 
and evaluating it at the point $(\tilde{\phi}, M_h^c)$, we get 
\begin{equation}
\label{diffNI}
\left. \frac{\partial^2 V_{\text{eff}}}{\partial \phi \ \partial \xi} \right|_{\tilde{\phi}, M_h^c} = 
- \left. \frac{\partial C}{\partial \phi} \right|_{\tilde{\phi}, M_h^c}  \left. \frac{\partial V_{\text{eff}}}{\partial \phi}  \right|_{\tilde{\phi}, M_h^c}   
- C(\tilde{\phi}, \xi) \left. \frac{\partial^2 V_{\text{eff}}}{\partial \phi^2} \right|_{\tilde{\phi}, M_h^c} \, . 
\end{equation}
The first term in the rhs of \eq{diffNI} vanishes because of the stationary condition in \eq{absvacstab2}. 
Hence, we can substitute the third term in \eq{totaldeiffxiB}, by means of \eq{diffNI}, and get:
\begin{equation}
\label{totaldeiffxiBsub}
\left( \frac{\partial \tilde{\phi}}{\partial \xi} - C(\tilde{\phi}, \xi) \right)  \left. \frac{\partial^2 V_{\text{eff}}}{\partial \phi^2} \right|_{\tilde{\phi}, M_h^c} = 0 \, .
\end{equation}
Since the curvature at the extremum is in general different from zero, \eq{totaldeiffxiBsub} implies
\begin{equation}
\label{gaugedepextr}
\frac{\partial \tilde{\phi}}{\partial \xi} = C(\tilde{\phi}, \xi) \, .
\end{equation}
The same holds for any extremum of the effective potential, like e.g.~the maximum 
in \fig{Mchplot} or the electroweak minimum $\phi_{\rm{ew}}$. 
This latter fact should not actually come as a surprise. 
The explicit gauge dependence of the unrenormalized $\phi_{\rm{ew}}$ 
in the $R_\xi$ gauge was discussed for instance in \cite{Appelquist:1973ms} 
and in the case of the SM it can be found in \cite{Sirlin:1985ux}. 
A renormalized gauge-invariant $\phi_{\rm{ew}}$ can always be defined by subtracting 
the divergent and gauge-dependent contributions to $\phi_{\rm{ew}}$ 
at on-shell points in terms of physical quantities.

\subsection{Gauge dependence of the SM vacuum instability scale}

The SM vacuum instability scale is operatively defined as the field value $\phi = \Lambda$, 
for which the effective potential has the same depth of the electroweak minimum (see e.g.~\fig{Mchplot}). 
This is analytically expressed by 
\begin{equation}
\label{definstscale2}
V_{\text{eff}} (\Lambda; \xi) = V_{\text{eff}} (\phi_{\rm{ew}}; \xi) \, . 
\end{equation}
The rhs of \eq{definstscale2} is a gauge-independent quantity, since $\phi_{\rm{ew}}$ is by definition a minimum 
and we can apply the Nielsen identity. 
Hence, by solving \eq{definstscale2}, one has in general 
$\Lambda = \Lambda (\xi)$.  
In particular, by taking the total differential of \eq{definstscale2} 
with respect to $\xi$, we get 
\begin{equation}
\label{totdiffdefinstscale2}
\left. \frac{\partial V_{\text{eff}}}{\partial \phi} \right|_{\Lambda} \frac{\partial \Lambda}{\partial \xi} 
+ \left. \frac{\partial V_{\text{eff}}}{\partial \xi} \right|_{\Lambda} = 0 \, . 
\end{equation}
By using the Nielsen identity, we can substitute back the second term in \eq{totdiffdefinstscale2}, thus obtaining 
\begin{equation}
\label{totdiffdefinstscale2bis}
\left( \frac{\partial \Lambda}{\partial \xi} - C(\Lambda, \xi) \right)  \left. \frac{\partial V_{\text{eff}}}{\partial \phi} \right|_{\Lambda} = 0 \, .
\end{equation}
Since, in general, $\Lambda$ is not an extremum of the effective potential, \eq{totdiffdefinstscale2bis} yields
\begin{equation}
\label{totdiffdefinstscale2bisyields}
\frac{\partial \Lambda}{\partial \xi} = C(\Lambda, \xi) \, .
\end{equation}

\clearpage

\section{Numerical analysis}
\label{gaugedepSMinst}

\begin{figure}[htb]
\centering
  \begin{tabular}{@{}cccc@{}}
    \includegraphics[width=.48\textwidth]{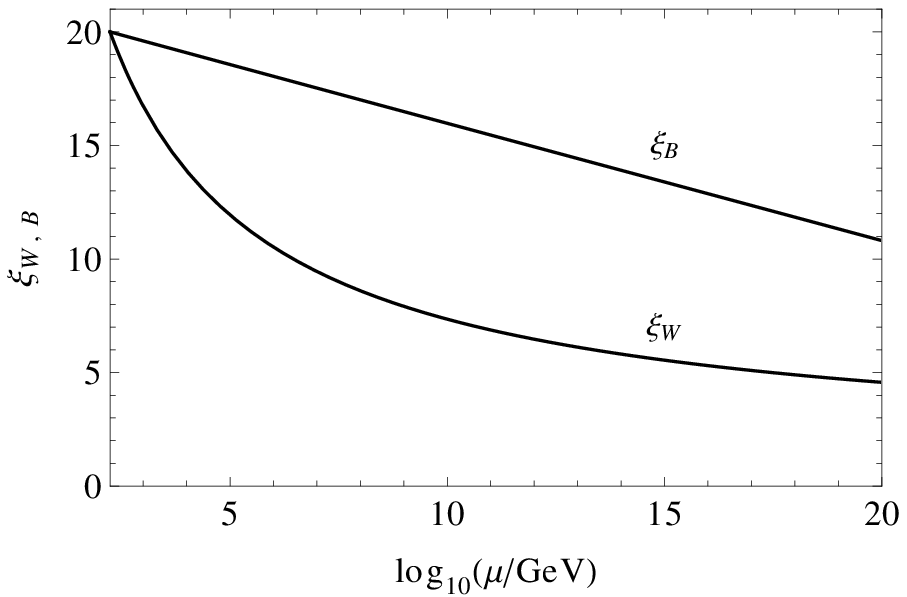} &
    \includegraphics[width=.48\textwidth]{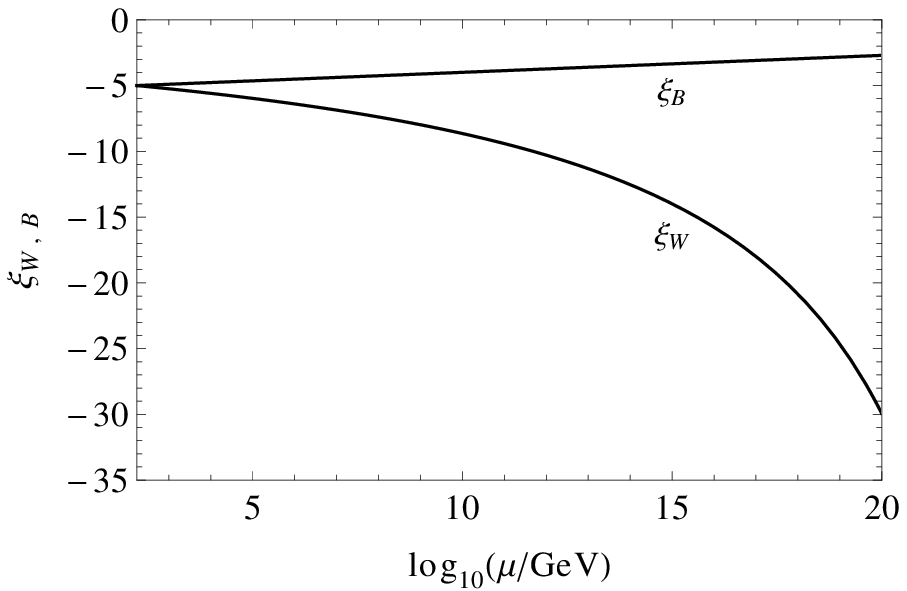} &
  \end{tabular}
  \caption{\label{runxi} 
  Two-loop running of the gauge-fixing parameters $\xi_W$ and $\xi_B$ in the Fermi gauge, 
  for different values of $\xi \equiv \xi_W (M_t) = \xi_B (M_t)$: $\xi = 20$ (left panel) and $\xi = -5$ (right panel).}
\end{figure}

In this Section we numerically estimate the gauge dependence of the SM vacuum
instability scale $\Lambda$. Let us first focus on the case of the Fermi gauge. 
Since in the SM $\Lambda \gg \phi_{\rm{ew}}$, the condition in \eq{definstscale2} is well approximated by 
(see also \eq{Veffapprox})  
\begin{equation}
\label{instLambda}
\lambda_{\rm{eff}} (\Lambda) = 0 \, , 
\end{equation}
up to corrections of ${\cal O}(\phi_{\rm{ew}}^2 / \Lambda^2)$ . 
For the onset of the RG running, we choose $\mu(0) = M_t$ (hence $\mu(t) = M_t e^t$), 
where $M_t = 173.35 \ \rm{GeV}$ is the pole 
mass of the top quark and we consider the central values of the SM parameters 
taken from \cite{Buttazzo:2013uya}:\footnote{Notice that these values are extracted from 
experimental data with two-loop accuracy. However, we will not perform a NNLO analysis, 
since the issue of the gauge dependence of the instability scale already arises at the NLO level.}
\begin{align}
\label{lamMt}
\lambda (M_t) &= 0.12710 \, , \\
\label{ytMt}
y_t (M_t) &= 0.93697 \, , \\
\label{g3Mt}
g_3 (M_t) &= 1.1666 \, , \\
\label{gMt}
g (M_t) &= 0.6483 \, , \\
\label{gpMt}
g' (M_t) &= 0.3587 \, . 
\end{align}
In order to resum possible large logs in \eq{lameffapprox} due to the growth of the anomalous dimension, 
we make the scale choice
\begin{equation}
\label{scalechoice}
\mu(\overline{t}) = e^{\Gamma(\overline{t})} \phi \, ,
\end{equation}
which implicitly defines $\overline{t}$ as a function of $\phi$. 
Then the effective quartic coupling can be written as
\begin{equation}
\label{lameffapproxscalechoice}
\lambda_{\rm{eff}}(\phi) = e^{4 \Gamma(\overline{t}(\phi))} \left[ \lambda(\overline{t}(\phi)) 
+ \frac{1}{(4\pi)^2} \sum_p N_p \kappa_p^2 (\overline{t}(\phi)) 
\left( \log \kappa_p(\overline{t}(\phi)) - C_p \right) \right] \, .
\end{equation}
Since the overall exponential factor in \eq{lameffapproxscalechoice} 
never changes the zeros of $\lambda_{\rm{eff}} (\phi)$, 
in order to find the instability scale, $\Lambda$, 
it is equivalent (and also numerically more convenient) 
to seek directly the zeros of $\lambda_{\rm{eff}} (\phi) e^{-4 \Gamma(\overline{t}(\phi))}$
in terms of the parameter $\overline{t}_\Lambda \equiv \overline{t}(\Lambda)$, 
defined by\footnote{It may actually happen that $\lambda$ turns negative before approaching the instability scale. 
In such a case, $\log \kappa_p$ develops an imaginary part for $p=h,A^{\pm},B^{\pm}$ 
(see \Table{tab:pvaluesFermi}). Though the imaginary part of the effective potential might have 
an interpretation in terms of a decay rate of an unstable state \cite{Weinberg:1987vp}, 
the role of such an imaginary component in the determination of the instability scale is not clear. 
Hence, we pragmatically require only the real part of \eq{zerossimplified} 
to be zero and notice that this problem has nothing to do with the issue of the gauge dependence, 
since it occurs also in the standard analysis in the Landau gauge.}
\begin{equation}
\label{zerossimplified}
\lambda(\overline{t}_\Lambda) + \frac{1}{(4\pi)^2} \sum_p N_p \kappa_p^2 (\overline{t}_\Lambda) 
\left( \log \kappa_p(\overline{t}_\Lambda) - C_p \right) = 0 \, ,
\end{equation}
and then relate it to the instability scale by inverting \eq{scalechoice}  
\begin{equation}
\label{explicitlysolv}
\Lambda = \mu(\overline{t}_\Lambda) e^{-\Gamma(\overline{t}_\Lambda)} = M_t e^{\overline{t}_\Lambda - \Gamma(\overline{t}_\Lambda)} \, ,
\end{equation}
where we recall the definition (see \eq{Gammat}) 
\begin{equation}
\label{GammatLambda}
\Gamma(\overline{t}_\Lambda) = - \int_0^{\overline{t}_\Lambda} \gamma(t) \, dt \, .
\end{equation}
Before discussing in more detail the gauge dependence of $\Lambda$, 
let us turn to the issue of the UV behaviour of the gauge fixing parameters 
$\xi_W$ and $\xi_B$ for the Fermi gauge. 
Their RGEs are collected in \app{RGEapp} and can 
be easily integrated at one loop (see \app{UVbehaviorxiBW}). 
While the running of the Abelian gauge-fixing parameter $\xi_B$  
is very simple ($\xi_B g'^2$ is actually constant under the RG flow, as a 
consequence of a Ward identity) 
two peculiar RG behaviours can be identified for $\xi_W$. 
For $\xi_W (M_t) \gg \frac{1}{6}$ one has a quasi-fixed point in the UV (cf.~left panel in \fig{runxi}),  
while, for $\xi_W (M_t) < 0$, the running can easily generate a Landau pole (cf.~right panel in \fig{runxi}). 

\begin{figure}[htb]
\centering
  \begin{tabular}{@{}cccc@{}}
    \includegraphics[width=.48\textwidth]{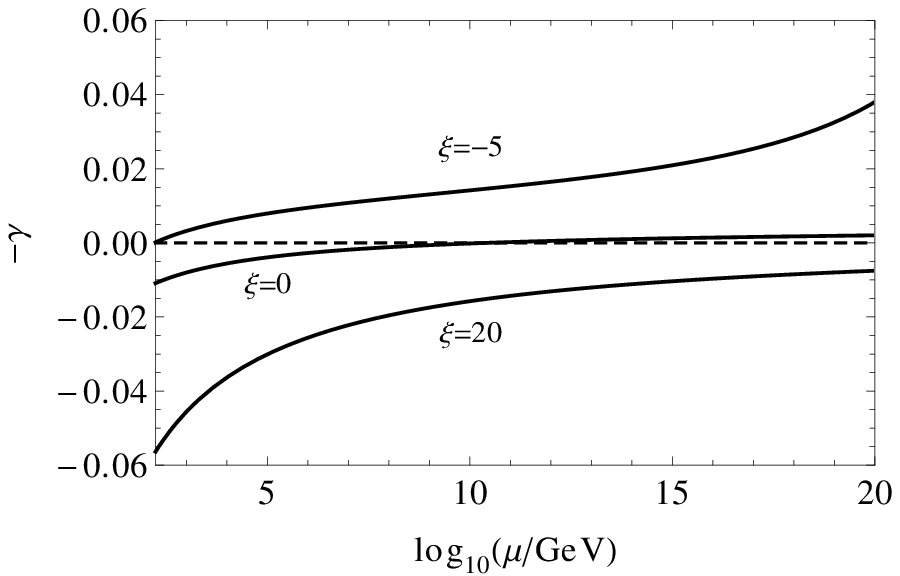} &
    \includegraphics[width=.48\textwidth]{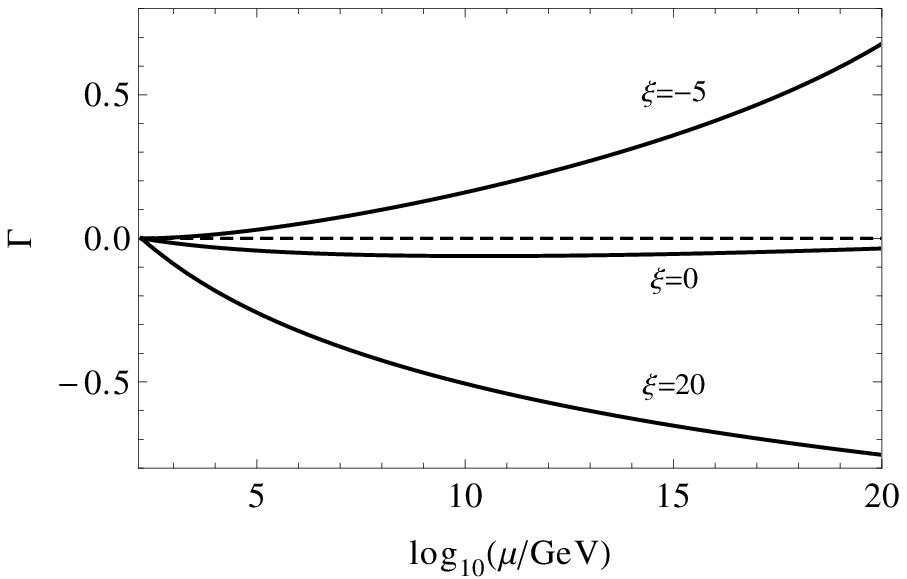} &
  \end{tabular}
  \caption{\label{rungGammaxi0205m} 
  Two-loop running of $-\gamma$ (left panel) and $\Gamma$ (right panel) for different 
  values of $\xi \equiv \xi_W (M_t) = \xi_B (M_t)$.}
\end{figure}

The gauge dependence of $\Lambda$ (cf.~\eq{explicitlysolv})
comes both from $t_\Lambda$ and $\Gamma (t_\Lambda)$. 
The former is due to the couplings $\kappa_p$, 
when $p$ runs over $A^{\pm}$ and $B^{\pm}$ (cf.~\eq{zerossimplified} and \Table{tab:pvaluesFermi}), 
while the latter is because of the gauge dependence of the anomalous dimension. 
The running of the anomalous dimension and its integral, $\Gamma$,  
are shown in \fig{rungGammaxi0205m} 
for three different initial values of $\xi \equiv \xi_B (M_t) = \xi_W (M_t)$.

From the right panel in \fig{rungGammaxi0205m} one can see that if $\abs{\xi}$ is large enough, 
$\Gamma$ can easily be of $\mathcal{O} (1)$ at intermediate scales below the Planck mass.  
This justifies the choice of scale done in \eq{scalechoice}, which resums the 
potentially large logs in \eq{lameffapprox}.

\begin{figure}[h]
\centering
\includegraphics[angle=0,width=11cm]{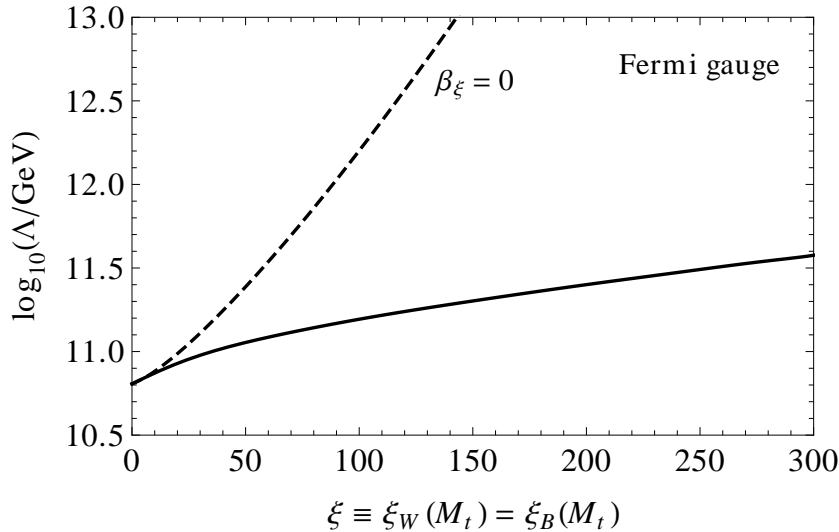}
\caption{\label{instscalevsxi} 
Instability scale as a function of $\xi \equiv \xi_W (M_t) = \xi_B (M_t)$ for the Fermi gauge.
The dashed line corresponds to the case where the gauge-fixing parameters are not run.  
The full line encodes the resummation of the next-to-leading logs by means of two-loop RGEs.  
}
\end{figure}

The gauge dependence of the instability scale is shown in
\fig{instscalevsxi}. For simplicity, we set $\xi_W (M_t) = \xi_B (M_t) \equiv
\xi$. In addition, we employ two-loop RGEs for all the parameters in
\eq{zerossimplified} and \eq{GammatLambda} that determine $\Lambda$. The
higher-order RGEs allow us to resum the leading and next-to-leading
logarithms implicitly contained in \eq{explicitlysolv}. For illustration, we 
depict with a dashed line in \fig{instscalevsxi} the gauge dependence of the
instability scale obtained without running the gauge-fixing parameters ($\beta_{\xi} = 0$ case).
As it can be read from the figure the difference between the resummed (full line)
and not resummed one (dashed line) amounts to more than three orders of
magnitude.  However, even after performing the resummation, the instability
scale in the Fermi gauge increases by almost an order of magnitude when the gauge-fixing parameters 
are varied in the interval $[0,300]$. Let us also mention that by varying the SM parameters within their experimental uncertainties (e.g.~for a lower top mass) the gauge dependence of the scale $\Lambda$ 
is always found to be of about one order of magnitude. 

Another important aspect for the analysis of the gauge dependence of $\Lambda$
is the determination of the perturbativity domain of the gauge fixing
parameters $\xi_{W,B}$. For instance, for the gauge-fixing parameter $\xi_W$ one can require that the 
two-loop correction to its beta function is smaller than the one-loop 
contribution, thus obtaining (cf.~\eq{RGE2lxiW} in \app{RGEapp}):
\begin{equation}
\label{pertdomxiW} 
\left| \frac{\xi_W^2 \alpha_2^2}{(4 \pi)^2} \right| < \left| \frac{\xi_W \alpha_2}{4 \pi} \right| \, , 
\end{equation}
which sets the absolute upper bound  
\begin{equation}
\label{pertdomxiWabsolute} 
 |\xi_W| < \frac{4 \pi}{\alpha_2} \, .  
\end{equation}
Taking $\alpha_2 (M_t) \approx 0.033$,\footnote{For $\alpha_2 (\mu > M_t)$ the bound becomes 
less stringent, due to the asymptotic freedom of $\alpha_2$ in the SM.}
one gets $|\xi_W (M_t)| < 376$.   
Notice, however, that this estimate does not take into account the running of $\xi_W$. 
For $\xi_W (M_t) \lesssim - 5$ a Landau pole can be developed before the Planck scale 
(cf.~right panel in \fig{runxi}), and perturbation theory starts soon to break down. This is why we 
do not show the negative branch of the plot in \fig{instscalevsxi}. 
On the contrary, the running behaviour for $\xi \gg 0$ is smoother, 
with a quasi-fixed point in the UV for $\xi_W$ (cf.~left panel in \fig{runxi}). 
By studying the evolution of the gauge-dependent anomalous dimension at one, two and three loops 
we verified, for instance, that $\xi \approx 300$ is still in the perturbative regime. 
Nonetheless, for a more solid statement about the perturbative domain of $\xi$, 
one should inspect the gauge dependent two-loop effective potential, whose calculation 
goes beyond the scope of the present paper and it is postponed for a future work. 
One can imagine, however, that a similar condition 
as in \eq{pertdomxiW} will be at play, since the gauge-fixing parameters are always associated 
with the square of the gauge couplings, both in the propagators and in the vertices of the theory. 

Finally, for a comprehensive analysis one should also vary the gauge-fixing
condition itself. In \app{BCKGgaugefull} we report on the calculation of the 
SM one-loop effective potential in a background $R_\xi$ gauge. 
A numerical study, similar to the one presented in this Section, 
shows that 
the instability scale decreases 
by another order of magnitude  
when the
gauge-fixing parameters are varied 
in their perturbative domain. 
Such a qualitatively different behaviour in the background $R_\xi$ gauge 
can be understood by noticing the sign flip (with respect to the case of the Fermi gauge) 
in the contribution of the gauge-fixing parameters to the one-loop anomalous dimension 
of $\phi$ in \eq{RGE2lphiBCKD}.
We can thus conclude that the gauge dependence of the instability scale 
materializes in a variation of about two orders of magnitude, depending on
the choice of the gauge condition and of the gauge-fixing parameters. 
This strengthens our statement that the instability scale $\Lambda$ as defined in
\eq{instLambda} should not be interpreted as a physical quantity. 

\clearpage

\section{Discussion and conclusions}
\label{conclusions}

Once a calculable UV completion of the SM is specified (for instance, the SM itself extrapolated at extremely high 
energies\footnote{Under the assumption that Planck-scale physics decouples from the SM even at energies beyond the Planck mass 
and that the Laundau pole of the hypercharge does not pose any conceptual problem.}) 
the fate of the electroweak vacuum, whether it is absolutely stable or not, 
is a physical statement which does not depend on the choice of the gauge. This is equivalent to say that the critical 
Higgs boson mass (or, in general, the critical values of the SM parameters) distinguishing between the stable and unstable phase of the SM 
is a gauge-independent quantity, as we formally proved in \sect{gaugeindepMHc}.
In this respect, it is worth to recall that the tunnelling probability of the electroweak vacuum is 
formally gauge independent as well \cite{Einhorn:1980ik,Metaxas:1995ab,Isidori:2001bm}. 

On the other hand, the absolute stability condition is sometimes formulated by requiring that the 
electroweak minimum, $\phi_{\rm{ew}}$, is the global minimum of the effective
potential over the range of validity of the SM  
\begin{equation}
\label{vacstabcond}
V_{\rm{eff}} (\phi_{\rm{ew}}) < V_{\rm{eff}} (\phi) \quad \rm{for} \quad  \phi < \Lambda_{\rm{SM}} \, , 
\end{equation}
where $\Lambda_{\rm{SM}}$ is a physical threshold (e.g.~the Planck scale). 
Above this scale new physics is supposed to alter the shape of the effective potential. 
However, since $V_{\rm{eff}} (\phi)$ is gauge dependent 
(unless $\phi$ is an extremum), 
the condition in \eq{vacstabcond} is clearly gauge dependent too. \\
From a low-energy point of view, it is a relevant question to seek 
a connection between the instability scale, $\Lambda$, and the scale of new 
physics, $\Lambda_{\rm{SM}}$. The latter being, of course, of utmost 
importance for experiments. 
The irreducible gauge dependence of $\Lambda$, however, makes its
identification with $\Lambda_{\rm{SM}}$ ambiguous,  
since we are not comparing two physical quantities.

Though the gauge dependence of $\Lambda$ amounts to about one order of magnitude in the case of the Fermi gauge (cf.~\fig{instscalevsxi}), 
this result \emph{cannot} be used to give an absolute upper bound on the gauge dependence of $\Lambda$. 
The reason is that, on one hand, different gauge-fixing schemes generally lead to different results 
(as, for instance, in the case of the background $R_\xi$ gauge discussed in \app{BCKGgaugefull}) 
and, on the other hand, 
we cannot say much beyond perturbation theory. Notice, indeed, that there is no physical principle 
that restricts the range of the gauge-fixing parameters. 
Hence, we rather stick to the conclusion that 
$\Lambda_{\rm{SM}}$ is a model dependent parameter which cannot be 
determined by just extrapolating the SM parameters at high 
energies.\footnote{Even without considering the issue of the gauge dependence, 
the connection between $\Lambda$ and the maximum allowed value 
of the scale of new physics required to stabilize the electroweak vacuum 
is anyway not so direct, 
due to the presence of extra parameters (e.g.~couplings and masses) in any 
UV completion of the SM \cite{Hung:1995in,Casas:2000mn}.}

Let us finally recall that, given the central values of the SM parameters and assuming that 
new physics at e.g.~the Planck scale does not affect the tunnelling computation \cite{Branchina:2013jra}, 
the lifetime of the electroweak vacuum turns out to be much 
longer than the age of the universe \cite{Buttazzo:2013uya}. 
A metastable electroweak vacuum can comply with the data and new physics is
not necessarily implied. 
Hence, the problem of the gauge dependence of the SM vacuum instability scale 
and its connection with the scale of new physics might seem an academic one. 
However, this does not need to be necessarily the case. 
For instance, we would like to mention the recent measurement of the primordial tensor fluctuations 
in the cosmic microwave background 
by the 
BICEP2 collaboration \cite{Ade:2014xna} which suggests 
a high inflationary scale of about $10^{14}$ GeV. 
As pointed out in \cite{Espinosa:2007qp,Kobakhidze:2013tn,Fairbairn:2014zia,Enqvist:2014bua,Kobakhidze:2014xda,Hook:2014uia} 
the Higgs field might be subject to quantum 
fluctuations generated during the primordial stage of inflation 
which can easily destabilize the electroweak 
vacuum. In particular, since the quantity $\Lambda$ (or, more precisely, the field 
value where the effective potential reaches its maximum) 
enters in the calculation of the 
electroweak vacuum survival probability, 
its physical identification should be addressed with care.

\section*{Acknowledgments}
We thank Stefano Bertolini, Ramona Gr{\"o}ber and Marco Nardecchia for useful discussions.
This work was supported by the DFG through the SFB/TR 9 ``Computational Particle Physic''.

\appendix

\section{Renormalization group equations}
\label{RGEapp}

In terms of the parameters 
$\alpha_1 = \frac{5}{3} \frac{g'^2}{4 \pi}$, \
$\alpha_2 = \frac{g^2}{4 \pi}$, 
$\alpha_3 = \frac{g_3^2}{4 \pi}$, 
$\alpha_t = \frac{y_t^2}{4 \pi}$ and
$\alpha_\lambda = \frac{\lambda}{4 \pi}$,
the two-loop RGEs used in the numerical analysis for the case of the Fermi gauge are 
\cite{Mihaila:2012pz,Bednyakov:2012rb,Chetyrkin:2013wya,Bednyakov:2013eba}
\begin{align}
\label{RGE2lalpha1}
\mu^2 \frac{d}{d \mu^2} \frac{\alpha_1}{\pi} & = 
\frac{41}{40} \frac{\alpha_1^2}{\pi^2}  
+\frac{199}{800} \frac{\alpha_1^3}{\pi ^3}
+\frac{27}{160} \frac{\alpha_1^2}{\pi^2}\frac{\alpha_2}{\pi} 
+\frac{11}{20}\frac{\alpha_1^2}{\pi^2} \frac{\alpha_3}{\pi} 
-\frac{17}{160}\frac{\alpha_1^2}{\pi^2}\frac{\alpha_t}{\pi}
\, , \\
\label{RGE2lalpha2}
\mu^2 \frac{d}{d \mu^2} \frac{\alpha_2}{\pi} & = 
- \frac{19}{24} \frac{\alpha_2^2}{\pi^2} 
+ \frac{9 }{160}\frac{\alpha_1}{\pi} \frac{\alpha_2^2}{\pi^2}
+\frac{35 }{96}\frac{\alpha_2^3}{\pi^3}
+\frac{3 }{4}\frac{\alpha_2^2}{\pi^2}\frac{\alpha_3}{\pi}
-\frac{3 }{32} \frac{\alpha_2^2}{\pi^2} \frac{\alpha_t}{\pi}
\, , \\
\label{RGE2lalpha3}
\mu^2 \frac{d}{d \mu^2} \frac{\alpha_3}{\pi} & = 
- \frac{7}{4} \frac{\alpha_3^2}{\pi^2} 
+ \frac{11 }{160} \frac{\alpha_1}{\pi} \frac{\alpha_3^2}{\pi^2}
+\frac{9 }{32} \frac{\alpha_2}{\pi} \frac{\alpha_3^2}{\pi^2}
-\frac{13}{8} \frac{\alpha_3^3}{\pi^3}
-\frac{1}{8} \frac{\alpha_3^2}{\pi^2} \frac{\alpha_t}{\pi} 
\, , \\
\label{RGE2lalphat}
\mu^2 \frac{d}{d \mu^2} \frac{\alpha_t}{\pi} & = \frac{\alpha_t}{\pi} 
\left( \frac{9}{8} \frac{\alpha_t}{\pi} - \frac{17}{80} \frac{\alpha_1}{\pi} - \frac{9}{16} \frac{\alpha_2}{\pi} 
- 2 \frac{\alpha_3}{\pi} \right) 
+ \frac{\alpha_t}{\pi} \left(
\frac{3}{8} \frac{\alpha _{\lambda }^2}{\pi^2} 
-\frac{3}{4} \frac{\alpha _{\lambda }}{\pi} \frac{\alpha _t}{\pi} 
-\frac{3}{4} \frac{\alpha _t^2}{\pi^2}  
+\frac{393 }{1280} \frac{\alpha _1}{\pi} \frac{\alpha _t}{\pi}
\right. \nonumber  \\
& \left.  
+\frac{225}{256} \frac{\alpha _2}{\pi} \frac{\alpha_t}{\pi}
+\frac{9}{4} \frac{\alpha _3}{\pi} \frac{\alpha _t}{\pi} 
+\frac{1187}{9600} \frac{\alpha _1^2}{\pi^2}
-\frac{23}{64} \frac{\alpha_2^2}{\pi^2} 
-\frac{27}{4} \frac{\alpha _3^2}{\pi^2}  
-\frac{9}{320} \frac{\alpha _1}{\pi} \frac{\alpha _2}{\pi} 
+\frac{19}{240} \frac{\alpha _1}{\pi} \frac{\alpha _3}{\pi}
\right. \nonumber \\
& \left.
+\frac{9}{16} \frac{\alpha _2}{\pi} \frac{\alpha _3}{\pi} \right)
\, ,  \\ 
\label{RGE2lalphalam}
\mu^2 \frac{d}{d \mu^2} \frac{\alpha_\lambda}{\pi} & = 
\frac{27}{1600} \frac{\alpha_1^2}{\pi^2} 
+ \frac{9}{160} \frac{\alpha_1}{\pi} \frac{\alpha_2}{\pi} 
+ \frac{9}{64} \frac{\alpha_2^2}{\pi^2} 
- \frac{3}{4} \frac{\alpha_t^2}{\pi^2} 
+ \frac{\alpha_\lambda}{\pi} \left( 
- \frac{9}{40} \frac{\alpha_1}{\pi}
- \frac{9}{8} \frac{\alpha_2}{\pi}
+ \frac{3}{2} \frac{\alpha_t}{\pi}
+ 3 \frac{\alpha_\lambda}{\pi}
\right) \\
& 
-\frac{3411}{64000} \frac{\alpha_1^3}{\pi^3}
-\frac{1677}{12800} \frac{\alpha_1^2}{\pi^2} \frac{\alpha_2}{\pi}
-\frac{171}{3200} \frac{\alpha_1^2}{\pi^2} \frac{\alpha_t}{\pi}
+\frac{1887}{6400} \frac{\alpha_1^2}{\pi^2} \frac{\alpha_\lambda}{\pi} 
-\frac{289}{2560} \frac{\alpha_1}{\pi} \frac{\alpha_2^2}{\pi^2}
+\frac{63}{320} \frac{\alpha_1}{\pi} \frac{\alpha_2}{\pi} \frac{\alpha_t}{\pi} \nonumber \\
& +\frac{117}{640} \frac{\alpha_1}{\pi} \frac{\alpha_2}{\pi} \frac{\alpha_\lambda}{\pi} 
-\frac{1}{20} \frac{\alpha_1}{\pi}  \frac{\alpha_t^2}{\pi^2}
+\frac{17}{64} \frac{\alpha_1}{\pi} \frac{\alpha_t}{\pi} \frac{\alpha_\lambda}{\pi} 
+\frac{27}{40} \frac{\alpha_1}{\pi}  \frac{\alpha_\lambda^2}{\pi^2} 
+\frac{305}{512} \frac{\alpha_2^3}{\pi^3}
-\frac{9}{128} \frac{\alpha_2^2}{\pi^2}  \frac{\alpha_t}{\pi}
-\frac{73}{256} \frac{\alpha_2^2}{\pi^2} \frac{\alpha_\lambda}{\pi} \nonumber \\
& +\frac{45}{64} \frac{\alpha_2}{\pi} \frac{\alpha_t}{\pi}  \frac{\alpha_\lambda}{\pi} 
+\frac{27}{8} \frac{\alpha_2}{\pi} \frac{\alpha_\lambda^2}{\pi}
-\frac{\alpha_3}{\pi} \frac{\alpha_t^2}{\pi^2}
+\frac{5}{2} \frac{\alpha_3}{\pi}  \frac{\alpha_t}{\pi}  \frac{\alpha_\lambda}{\pi}  
+\frac{15}{16} \frac{\alpha_t^3}{\pi^3}
-\frac{3}{32} \frac{\alpha_t^2}{\pi}  \frac{\alpha_\lambda}{\pi} 
-\frac{9}{2} \frac{\alpha_t}{\pi}  \frac{\alpha_\lambda^2}{\pi^2}
-\frac{39}{4} \frac{\alpha_\lambda^3}{\pi^3}
\, , \nonumber \\
\label{RGE2lxiB}
\mu^2 \frac{d}{d\mu^2} \frac{\xi_B}{\pi} &= \frac{\xi_B}{\pi} \left( -\frac{41}{40} \frac{\alpha_1}{\pi} \right)
+ \frac{\xi_B}{\pi} \left( -\frac{199}{800} \frac{\alpha_1^2}{\pi^2} -\frac{27}{160} \frac{\alpha_1}{\pi} \frac{\alpha_2}{\pi} 
 -\frac{11}{20} \frac{\alpha_1}{\pi} \frac{\alpha_3}{\pi} 
 +\frac{17}{160} \frac{\alpha_1}{\pi} \frac{\alpha_t}{\pi} \right) \, , \\
\label{RGE2lxiW}
\mu^2 \frac{d}{d\mu^2} \frac{\xi_W}{\pi} &= \frac{\xi_W}{\pi} \left( \frac{1}{24} \frac{\alpha_2}{\pi} - \frac{1}{4} 
\frac{\xi_W \alpha_2}{\pi} \right) \nonumber \\
&
+ \frac{\xi_W}{\pi} \left( -\frac{9}{160} \frac{\alpha_1}{\pi} \frac{\alpha_2}{\pi} 
-\frac{43}{64} \frac{\alpha_2^2}{\pi^2}
-\frac{3}{4} \frac{\alpha_2}{\pi} \frac{\alpha_3}{\pi} 
+\frac{3}{32} \frac{\alpha_2}{\pi} \frac{\alpha_t}{\pi}
-\frac{11}{32} \frac{\alpha_2}{\pi} \frac{\xi_W \alpha_2}{\pi} 
-\frac{1}{16} \frac{\xi_W^2 \alpha_2^2}{\pi^2}
\right) \, , 
\\
\label{RGE2lphi}
\mu \frac{d}{d\mu} \phi &= - \phi 
\left( 
-\frac{9}{80} \frac{\alpha_1}{\pi} 
-\frac{9}{16} \frac{\alpha_2}{\pi}
+\frac{3}{4} \frac{\alpha_t}{\pi}
+\frac{3}{80} \frac{\xi_B \alpha_1}{\pi}
+\frac{3}{16} \frac{\xi_W \alpha_2}{\pi}
\right) 
 \nonumber \\ 
& - \phi 
\left( 
\frac{3 }{8} \frac{\alpha _{\lambda }^2}{\pi^2}
+\frac{1293 }{12800} \frac{\alpha _1^2}{\pi^2}
+\frac{27 }{1280} \frac{\alpha _2}{\pi} \frac{\alpha_1}{\pi}
-\frac{271 }{512} \frac{\alpha _2^2}{\pi^2}
+\frac{17 }{128} \frac{\alpha _1}{\pi} \frac{\alpha _t}{\pi}
-\frac{27 }{64} \frac{\alpha_t^2}{\pi^2}
+\frac{45 }{128} \frac{\alpha _2}{\pi} \frac{\alpha _t}{\pi} 
\right. \nonumber \\
& \left.  
+\frac{5 }{4} \frac{\alpha _3}{\pi} \frac{\alpha _t}{\pi}
+\frac{3}{16} \frac{\xi _W \alpha _2^2}{\pi^2}
+\frac{3}{128} \frac{\xi _W^2 \alpha_2^2}{\pi^2}
\right)
\, . 
\end{align}
In the case of the background $R_\xi$ gauge (see \app{BCKGgaugefull}), 
the one-loop running of the field $\phi$ is found to be 
\begin{equation}
\label{RGE2lphiBCKD}
\mu \frac{d}{d\mu} \phi = - \phi 
\left( 
-\frac{9}{80} \frac{\alpha_1}{\pi} 
-\frac{9}{16} \frac{\alpha_2}{\pi}
+\frac{3}{4} \frac{\alpha_t}{\pi}
-\frac{3}{80} \frac{\overline{\xi}_B \alpha_1}{\pi}
-\frac{3}{16} \frac{\overline{\xi}_W \alpha_2}{\pi}
\right) \, .
\end{equation}
Notice that, by perturbatively expanding the RGE satisfied by the effective potential 
in \eq{RGE} at the first non-trivial order, the gauge-dependent parts of the one-loop 
anomalous dimension can be extracted from the $\mu$-dependent terms of 
$V^{(1)}_{\rm{eff}}$, which provides a non-trivial check of the calculation.  


\subsection{On the UV behaviour of $\xi_B$ and $\xi_W$} 
\label{UVbehaviorxiBW}

To better understand the running properties of $\xi_B$ and $\xi_W$, 
it turns out to be useful to solve analytically \eqs{RGE2lalpha1}{RGE2lalpha2} and \eqs{RGE2lxiB}{RGE2lxiW}. 
At one loop we have
\begin{align}
\label{runalpha1anal}
\alpha_1 (\mu) &= \frac{\alpha_1 (M_t)}{1 - \frac{41}{20} \frac{\alpha_1(M_t)}{\pi} \log \frac{\mu}{M_t}} \, , \\
\label{runalpha2anal}
\alpha_2 (\mu) &= \frac{\alpha_2 (M_t)}{1 + \frac{19}{12} \frac{\alpha_2(M_t)}{\pi} \log \frac{\mu}{M_t}} \, , \\
\label{runxiBanal}
\xi_B (\mu) &= \xi_B (M_t) \left( 1 - \frac{41}{20} \frac{\alpha_1(M_t)}{\pi} \log \frac{\mu}{M_t} \right) \, , \\
\label{runxiWanal}
\xi_W (\mu) &= \frac{\frac{\xi_W(M_t)}{1-6\xi_W(M_t)}
\left( 1 + \frac{19}{12} \frac{\alpha_2(M_t)}{\pi} \log \frac{\mu}{M_t} \right)^{\tfrac{1}{19}}}
{1 + 6 \frac{\xi_W(M_t)}{1-6\xi_W(M_t)}
\left( 1 + \frac{19}{12} \frac{\alpha_2(M_t)}{\pi} \log \frac{\mu}{M_t} \right)^{\tfrac{1}{19}}} \, . 
\end{align}
The main features of the system of equations above can be summarized as follows:
\begin{itemize}
\item From \eq{RGE2lalpha1} and \eq{RGE2lxiB} (or, equivalently, from \eq{runalpha1anal} and \eq{runxiBanal}) 
it follows that $\alpha_1 \xi_B$ is constant under the RG flow.
This property is true at all orders in perturbation theory and is a consequence of the Ward identity 
$Z^B_3 Z_{\alpha_1} = 1$, where $Z^B_3$ and $Z_{\alpha_1}$ are respectively the 
hypercharge wavefunction and vertex renormalization constants. 
\item The values $\xi_B = 0$ and $\xi_W = 0$ are fixed points of the RG flow. This property is true at all orders in perturbation theory 
and guarantees that in the Landau gauge $\xi_B \neq 0$ and $\xi_W \neq 0$ are not radiatively generated. 
\item The value $\xi_W = \frac{1}{6}$ is a fixed point of the RG flow at one loop (cf.~\eq{RGE2lxiW}). 
However, such a property does not hold anymore 
at higher orders.  
\item 
For $\xi_W (M_t) \gg \frac{1}{6}$ and $\mu \gg M_t$, \eq{runxiWanal} reaches the asymptotic value 
\begin{equation}
\label{runxiWapprox} 
 \xi_W (\mu) \approx \frac{-\frac{1}{6}
\left( 1 + \frac{19}{12} \frac{\alpha_2(M_t)}{\pi} \log \frac{\mu}{M_t} \right)^{\tfrac{1}{19}}}
{1 - \left( 1 + \frac{19}{12} \frac{\alpha_2(M_t)}{\pi} \log \frac{\mu}{M_t} \right)^{\tfrac{1}{19}}} \, ,
\end{equation}
which is independent from the initial condition $\xi_W (M_t)$ and always $> 0$. A typical RG 
solution in such a case is plotted in the left panel of \fig{runxi}. 
\item For $\xi_W (M_t) < 0$, \eq{runxiWanal} can develop a Landau pole. See e.g.~the right panel in \fig{runxi}.
\end{itemize}

\section{Background $R_\xi$ gauge}
\label{BCKGgaugefull}

In this appendix we consider the calculation of the SM one-loop effective potential 
in a generalization of the 
renormalizable 't Hooft gauge (see e.g.~\cite{Bohm:1986rj}) where the Higgs vacuum expectation value (vev) is promoted 
to the background field $\phi$. This is obtained by considering the following Lagrangian density 
\begin{equation}
\label{gflagBKGD}
\mathcal{L}^{\rm{BKGD}}_{\rm{g.f.}}  = 
-\frac{1}{2} \left[ 2 \bar{F}^+ \bar{F}^- + \left(\bar{F}^3\right)^2 + \left(\bar{F}^B\right)^2 \right] \, , 
\end{equation}
where the gauge-fixing functionals are defined as 
\begin{align}
\label{FpmdefBKGD}
\bar{F}^{\pm} &= \bar{\xi}_{1,W}^{-1/2} \partial^\mu W^\pm_\mu \mp i \bar{\xi}_{2,W}^{1/2} \bar{m}_W \chi^{\pm} \, , \\  
\label{FZdefBKGD}
\bar{F}^{3} &= \bar{\xi}_{1,3}^{-1/2} \partial^\mu W^3_\mu - \bar{\xi}_{2,3}^{1/2} \bar{m}_W \chi^{3} \, , \\
\label{FgammadefBKGD}
\bar{F}^{B} &= \bar{\xi}_{1,B}^{-1/2} \partial^\mu B_\mu - \bar{\xi}_{2,B}^{1/2} \bar{m}_B \chi^{3} \, ,  
\end{align}
with $W^\pm_\mu$ and $\chi^{\pm}$ conforming to the standard definitions
\begin{align}
\label{defWpm}
W^\pm_\mu &= \tfrac{1}{\sqrt{2}} \left( W^1_\mu \mp i W^2_\mu \right) \, , \\
\label{defchipm}
\chi^{\pm} &= \tfrac{1}{\sqrt{2}} \left(\chi^1 \pm i \chi^2 \right) \, . 
\end{align}
In \eqs{FpmdefBKGD}{FgammadefBKGD}, $\bar{m}_W$ and $\bar{m}_B$ are background-field-dependent masses 
(see \eqs{defmwFermi}{defmbFermi}) and the gauge-fixing parameters 
$\bar{\xi}_{1,\alpha},\bar{\xi}_{2,\alpha}$ (for $\alpha= W,3,B$) 
are denoted differently, since they have a different renormalization constant already at one loop \cite{Bohm:1986rj}.  

As long as we are not interested in the running properties of $\bar{\xi}_{1,\alpha}$ and $\bar{\xi}_{2,\alpha}$, 
they can be chosen equal at a given renormalization scale. This simplifies the
calculation of the one-loop effective potential,  
since the mixed Goldstone--gauge boson propagators do not appear at tree level. 
In a first step, we set for simplicity $\bar{\xi}_{1,W} = \bar{\xi}_{2,W} = \bar{\xi}_{1,3} = \bar{\xi}_{2,3} \equiv \bar{\xi}_W$ 
and $\bar{\xi}_{1,B} = \bar{\xi}_{2,B} \equiv \bar{\xi}_B$. For the full result with general
gauge-fixing parameters 
we refer to \app{fullres}.\footnote{We are 
aware of a similar calculation in the background $R_\xi$ gauge 
where all the gauge-fixing parameters in \eqs{FpmdefBKGD}{FgammadefBKGD} are taken equal \cite{Patel:2011th}.} 

A new feature, with respect to the Fermi gauge, is the non-trivial contribution of the ghost fields, 
which must be taken into account by means of the compensating ghost Lagrangian 
associated to the gauge-fixing functionals in \eqs{FpmdefBKGD}{FZdefBKGD}
\begin{equation}
\label{FPlagBCKG}
\mathcal{L}_{\rm{ghost}}^{\rm{BKGD}}= \sum_{\alpha\beta} c^\dag_\alpha \frac{\delta \bar{F}^\alpha}{\delta \theta^\beta} c_\beta \, ,
\end{equation}
where $c_\alpha$, $c^\dag_\alpha$ ($\alpha = +, -, 3, B)$ are the Feddeev-Popov ghost fields 
and $\delta/\delta \theta^\beta$ denotes the derivative with respect to the
parameter of the gauge transformation.   
Following the definition of the covariant derivative in \eq{defcovder}, 
the quadratic part of the ghost Lagrangian is found to be 
\begin{align}
\label{FPlagBCKGquad}
\mathcal{L}_{\rm{ghost}}^{\rm{BKGD/quad}} &= 
c_+^\dag \left( -\bar{\xi}_{W}^{-1/2} \Box - \bar{\xi}_{W}^{1/2} \bar{m}_W^2 \right) c_+ 
+ c_-^\dag \left( -\bar{\xi}_{W}^{-1/2} \Box - \bar{\xi}_{W}^{1/2} \bar{m}_W^2 \right) c_- 
\nonumber \\
& + c_3^\dag \left( -\bar{\xi}_{W}^{-1/2} \Box - \bar{\xi}_{W}^{1/2} \bar{m}_W^2 \right) c_3 
+ c_B^\dag \left( -\bar{\xi}_{B}^{-1/2} \Box - \bar{\xi}_{B}^{1/2} \bar{m}_B^2 \right) c_B \nonumber \\
&
+ c_3^\dag \left( - \bar{\xi}_{W}^{1/2} \bar{m}_W \bar{m}_B \right) c_B 
+ c_B^\dag \left( - \bar{\xi}_{B}^{1/2} \bar{m}_W \bar{m}_B \right) c_3 
\, .
\end{align}
Correspondingly, the inverse propagator matrix of the ghost fields in momentum space is given by
\begin{equation}
\label{invpropghostBCKG}
i \tilde{\mathcal{D}}^{-1}_{\rm{ghost}} = 
\left(
\begin{array}{cccc}
\bar{\xi}_{W}^{-1/2} k^2 - \bar{\xi}_{W}^{1/2} \bar{m}_W^2 & 0 & 0 & 0 \\
0 & \bar{\xi}_{W}^{-1/2} k^2 - \bar{\xi}_{W}^{1/2} \bar{m}_W^2 & 0 & 0 \\
0 & 0 & \bar{\xi}_{W}^{-1/2} k^2 - \bar{\xi}_{W}^{1/2} \bar{m}_W^2 & - \bar{\xi}_{W}^{1/2} \bar{m}_W \bar{m}_B \\
0 & 0 & - \bar{\xi}_{B}^{1/2} \bar{m}_W \bar{m}_B  & \bar{\xi}_{B}^{-1/2} k^2 - \bar{\xi}_{B}^{1/2} \bar{m}_B^2   
\end{array}
\right) \, ,
\end{equation}
defined on the complex field vector basis, $c^T = \left( c_{+}, c_{-}, c_{3}, c_{B} \right)$. 
Then from \eq{invpropghostBCKG} one gets
\begin{equation}
\label{logdetDm1ghost}
\log\det i \tilde{\mathcal{D}}^{-1}_{\rm{ghost}} = 
2 \log \left( k^2 - \bar{\xi}_{W} \bar{m}_W^2 \right)
+ \log \left( k^2 - \bar{\xi}_{W} \bar{m}_W^2 - \bar{\xi}_{B} \bar{m}_B^2 \right) 
+ \ldots \, ,
\end{equation}
where the ellipses stand for $\phi$-independent terms. 
The rest of the calculation proceeds as in \sect{Fermigauge}, 
with only two differences: the absence of the Goldstone--gauge boson mixing term, $\bar{m}_{\text{mix}}$, 
and the presence of an extra, gauge-dependent, contribution to the Goldstone boson masses
\begin{equation}
\label{invpropchiBCKG}
i \tilde{\mathcal{D}}^{-1}_{\chi} = 
\left(
\begin{array}{ccc}
k^2 - \bar{m}_\chi^2 - \bar{\xi}_{W} \bar{m}_W^2 & 0 & 0 \\
0 & k^2 - \bar{m}_\chi^2 - \bar{\xi}_{W} \bar{m}_W^2 & 0 \\
0 & 0 &k^2 - \bar{m}_\chi^2  - \bar{\xi}_{W} \bar{m}_W^2 - \bar{\xi}_{B} \bar{m}_B^2
\end{array}
\right) \, .
\end{equation}
Including all the relevant degrees of freedom, 
the one-loop effective potential is given by (cf.~\eq{1loopEPclosed})
\begin{align}
\label{EP1loopSMdrexplBCKG}
V^{(1)}_{\rm{eff}} (\phi) |^{\rm{BKGD}} &= -\frac{i}{2} \mu^{2\epsilon} \int \frac{d^d k}{(2\pi)^d} 
\left[
- 12 \log \left( - k^2 + \bar{m}_t^2 \right) + (d-1) \left( 2 \log \left( -k^2 + \bar{m}_W^2 \right) 
\right. \right. \nonumber \\ 
& \left. + \log \left( -k^2 + \bar{m}_Z^2 \right) \right) + \log \left( k^2 - \bar{m}_h^2 \right) 
+ 2 \log \left( k^2 - \bar{m}_{\chi^+}^2 \right) 
+ \log \left( k^2 - \bar{m}_{\chi^0}^2  \right) 
\nonumber \\ 
& \left. - 2 \log \left( k^2 - \bar{m}_{c_W}^2 \right) 
- \log \left( k^2 - \bar{m}_{c_Z}^2 \right) + \, \text{$\phi$-independent} \right] \, ,
\end{align} 
where we defined the field-dependent masses:
\begin{align}
\label{mcWbar}
\bar{m}_{c_W}^2 &= \bar{\xi}_{W} \bar{m}_W^2 \, , \\
\label{mcZhibar}
\bar{m}_{c_Z}^2 &= \bar{\xi}_{W} \bar{m}_W^2 + \bar{\xi}_{B} \bar{m}_B^2 \, , \\
\label{mchiphibar}
\bar{m}_{\chi^+}^2 &= \bar{m}_\chi^2 + \bar{\xi}_{W} \bar{m}_W^2 \, , \\
\label{mchi0hibar}
\bar{m}_{\chi^0}^2 &= \bar{m}_\chi^2 + \bar{\xi}_{W} \bar{m}_W^2 + \bar{\xi}_{B} \bar{m}_B^2  \, .
\end{align}
By performing the integral in \eq{EP1loopSMdrexplFermi} and by expanding in $\epsilon$, we get 
\begin{align}
\label{1loopEPbareBCKG} 
& V^{(1)}_{\rm{eff}}|_{\rm{bare}}^{\rm{BKGD}} = \frac{1}{4 (4 \pi)^2} \left[ 
-12 \bar{m}_t^4 \left( \log\frac{\bar{m}_t^2}{\mu^2} - \frac{3}{2} - \Delta_\epsilon \right)
+6 \bar{m}_W^4 \left( \log\frac{\bar{m}_W^2}{\mu^2} - \frac{5}{6} - \Delta_\epsilon \right) \right. \\
& \left.+3 \bar{m}_Z^4 \left( \log\frac{\bar{m}_Z^2}{\mu^2} - \frac{5}{6} - \Delta_\epsilon \right) 
+\bar{m}_h^4 \left( \log\frac{\bar{m}_h^2}{\mu^2} - \frac{3}{2} - \Delta_\epsilon \right)
+2 \bar{m}_{\chi^+}^4 \left( \log\frac{\bar{m}_{\chi^+}^2}{\mu^2} - \frac{3}{2} - \Delta_\epsilon \right)  \right. \nonumber \\
& \left. +  \bar{m}_{\chi^0}^4 \left( \log\frac{\bar{m}_{\chi^0}^2}{\mu^2} - \frac{3}{2} - \Delta_\epsilon \right) 
-2 \bar{m}_{c_W}^4 \left( \log\frac{\bar{m}_{c_W}^2}{\mu^2} - \frac{3}{2} - \Delta_\epsilon \right)
-  \bar{m}_{c_Z}^4 \left( \log\frac{\bar{m}_{c_Z}^2}{\mu^2} - \frac{3}{2} - \Delta_\epsilon \right)
\right] \, , \nonumber 
\end{align}
whose divergent part is explicitly given by
\begin{align}
\label{1loopdivBCKG} 
V^{(1)}_{\rm{eff}}|_{\rm{bare-pole}}^{\rm{BKGD}} &= \frac{\Delta_\epsilon}{(4\pi)^2} 
\left[
-m^4
+ \left(3 \lambda + \frac{1}{8} \bar{\xi} _B g'^2+\frac{3}{8} \bar{\xi} _W g^2   \right) m^2 \phi^2 \right. \nonumber \\
& \left. +
   \left(
   -\frac{3}{64} g'^4
   -\frac{3}{32} g'^2 g^2
   -\frac{9}{64} g^4
   +\frac{3}{4} y_t^4
   -3 \lambda ^2
   -\frac{1}{8} \bar{\xi} _B g'^2 \lambda 
   -\frac{3}{8}  \bar{\xi} _W g^2  \lambda 
   \right) \phi^4
\right] \, .
\end{align}
Notice that the divergent structure of \eq{1loopdivBCKG} 
can be identified with that in \eq{1loopdivFermi} of the Fermi gauge,  
after the replacement $ \overline{\xi}_{W,B} \rightarrow - \xi_{W,B}$. 
Hence, in order to cancel the gauge-dependent poles in \eq{1loopdivBCKG}, the same 
substitution must be made in the field renormalization constant in \eq{ZphiFermi}, 
which implies
\begin{equation}
\label{ZphiBCKG} 
Z_{\phi}^{1/2}|^{\rm{BKGD}} = 1 + 
\frac{\Delta_\epsilon}{(4\pi)^2} 
\left( 
\frac{3}{8} g'^2 + \frac{9}{8} g^2 -\frac{3}{2} y_t^2 +\frac{1}{8} \overline{\xi}_B g'^2 + \frac{3}{8} \overline{\xi}_W g^2   
\right) \, . \\
\end{equation}
The renormalization constants of $m^2$ and $\lambda$ are gauge independent and 
hence are given by the expressions in \eqs{Zm2Fermi}{ZlamFermi}.

After the subtraction of all the poles due to the renormalization prescription, 
the one-loop contribution to the effective potential in the $\overline{\rm{MS}}$ scheme 
reads
\begin{align}
\label{1loopEPBCKG} 
V^{(1)}_{\rm{eff}} (\phi)|^{\rm{BKGD}} &= \frac{1}{4 (4 \pi)^2} \left[ 
-12 \bar{m}_t^4 \left( \log\frac{\bar{m}_t^2}{\mu^2} - \frac{3}{2} \right)
+6 \bar{m}_W^4 \left( \log\frac{\bar{m}_W^2}{\mu^2} - \frac{5}{6} \right)
+3 \bar{m}_Z^4 \left( \log\frac{\bar{m}_Z^2}{\mu^2} - \frac{5}{6} \right) \right. \nonumber \\
& \left. +\bar{m}_h^4 \left( \log\frac{\bar{m}_h^2}{\mu^2} - \frac{3}{2} \right)
+2 \bar{m}_{\chi^+}^4 \left( \log\frac{\bar{m}_{\chi^+}^2}{\mu^2} - \frac{3}{2} \right)
+  \bar{m}_{\chi^0}^4 \left( \log\frac{\bar{m}_{\chi^0}^2}{\mu^2} - \frac{3}{2} \right) \right. \nonumber \\
& \left. -2 \bar{m}_{c_W}^4 \left( \log\frac{\bar{m}_{c_W}^2}{\mu^2} - \frac{3}{2} \right)
-  \bar{m}_{c_Z}^4 \left( \log\frac{\bar{m}_{c_Z}^2}{\mu^2} - \frac{3}{2} \right)
\right] \, , 
\end{align}
where the definition of the $\phi$-dependent mass terms can be found in \eqs{mcWbar}{mchi0hibar} 
(see also \eqs{mhphiFermi}{mtphiFermi}).   
For $\overline{\xi}_W = \overline{\xi}_B = 0$, \eq{1loopEPBCKG} 
reproduces the standard one-loop result in the Landau gauge \cite{Coleman:1973jx}, 
while, for $\overline{\xi}_W = \overline{\xi}_B$, it reproduces the result of \cite{Patel:2011th}. 
Moreover, on the tree-level minimum, $\bar{m}_{\chi} = 0$, one has 
$\bar{m}_{\chi^+} = \bar{m}_{c_W}$ and $\bar{m}_{\chi^0} = \bar{m}_{c_Z}$, 
so that the gauge dependence drops from $V^{(1)}_{\rm{eff}}$. 

By expanding \eq{1loopEPBCKG} in the $\phi \gg m$ limit, 
one gets the RG improved $\lambda_{\rm{eff}}$ coupling defined in \eq{lameffapprox}, 
with the $p$-coefficients explicitly given in \Table{tab:pvaluesBCKG}.
\begin{table*}[h]
  \begin{center}  
      \begin{tabular}{|c|cccccccc|}
      \hline
        $p$ & $t$ & $W$ & $Z$ & $h$ & $\chi^+$ & $\chi^0$ & $c_W$ & $c_Z$ \\ 
        \hline
        $N_p$ & $-12$ & $6$ & $3$ & $1$ & $2$ & $1$ & $-2$ & $-1$ \\ 
        $C_p$ & $\frac{3}{2}$ & $\frac{5}{6}$ & $\frac{5}{6}$ & $\frac{3}{2}$ & $\frac{3}{2}$ & $\frac{3}{2}$ & $\frac{3}{2}$ & $\frac{3}{2}$ \\ 
        $\kappa_p$ & $\frac{y_t^2}{2}$ & $\frac{g^2}{4}$ & $\frac{g^2+g'^2}{4}$ & $3 \lambda$ 
        & $\lambda + \frac{\bar{\xi}_W g^2}{4}$ & $\lambda + \frac{\bar{\xi}_B g'^2}{4} + \frac{\bar{\xi}_W g^2}{4}$ 
        & $\frac{\bar{\xi}_W g^2}{4}$ & $\frac{\bar{\xi}_B g'^2}{4} + \frac{\bar{\xi}_W g^2}{4}$ \\ 
        \hline
        \end{tabular}
    \caption{\label{tab:pvaluesBCKG} The $p$-coefficients entering the expression of 
    $\lambda_{\rm{eff}}$ in \eq{lameffapprox} 
    for the background $R_\xi$ gauge. 
      }
  \end{center}
\end{table*}

\subsection{Full result}
\label{fullres}

The expression of the effective potential in the background $R_\xi$ gauge 
for a general set of gauge-fixing parameters 
$\bar{\xi}_{1,\alpha},\bar{\xi}_{2,\alpha}$ ($\alpha= W,3,B$) is found to be
\begin{align}
\label{1loopEPbareBCKGFull} 
V^{(1)}_{\rm{eff}}|^{\rm{BKGD}} &= \frac{1}{4 (4 \pi)^2} \left[ 
-12 \bar{m}_t^4 \left( \log\frac{\bar{m}_t^2}{\mu^2} - \frac{3}{2} \right)
+6 \bar{m}_W^4 \left( \log\frac{\bar{m}_W^2}{\mu^2} - \frac{5}{6}  \right) 
+3 \bar{m}_Z^4 \left( \log\frac{\bar{m}_Z^2}{\mu^2} - \frac{5}{6}  \right) \right. \nonumber \\
& \left. 
+\bar{m}_h^4 \left( \log\frac{\bar{m}_h^2}{\mu^2} - \frac{3}{2}  \right)
+2 \bar{m}_{A^+}^4 \left( \log\frac{\bar{m}_{A^+}^2}{\mu^2} - \frac{3}{2}  \right)  
+2 \bar{m}_{A^-}^4 \left( \log\frac{\bar{m}_{A^-}^2}{\mu^2} - \frac{3}{2}  \right) 
\right. \nonumber \\
&   
+ \bar{m}_{B^+}^4 \left( \log\frac{\bar{m}_{B^+}^2}{\mu^2} - \frac{3}{2}  \right)
+ \bar{m}_{B^-}^4 \left( \log\frac{\bar{m}_{B^-}^2}{\mu^2} - \frac{3}{2} \right) 
-4 \bar{m}_{c_W}^4 \left( \log\frac{\bar{m}_{c_W}^2}{\mu^2} - \frac{3}{2}  \right) \nonumber \\
& \left. - 2 \bar{m}_{c_Z}^4 \left( \log\frac{\bar{m}_{c_Z}^2}{\mu^2} - \frac{3}{2} \right)
\right] \, ,  
\end{align}
where we employed the $\phi$-dependent masses in \eqs{mhphiFermi}{mtphiFermi} and further defined 
\begin{align} 
\label{defmassCpmFull}
\bar{m}_{A^{\pm}}^2 &= \frac{1}{2} \left(  \bar{m}_\chi^2 + 2 \sqrt{\bar{\xi}_{1,W}\bar{\xi}_{2,W}} \bar{m}_W^2 
\pm \bar{m}_\chi \sqrt{ \bar{m}_\chi^2 - 4 \left(\bar{\xi}_{1,W} - \sqrt{\bar{\xi}_{1,W}\bar{\xi}_{2,W} } \right) \bar{m}_W^2} \right) \, , \\ 
\label{defmassDpmFull}
\bar{m}_{B^{\pm}}^2 &= \frac{1}{2} \left(  \bar{m}_\chi^2 
+ 2 \sqrt{\bar{\xi}_{1,3}\bar{\xi}_{2,3}} \bar{m}_W^2 + 2 \sqrt{\bar{\xi}_{1,B}\bar{\xi}_{2,B}} \bar{m}_B^2 \right. \nonumber \\
& \left. \pm \bar{m}_\chi \sqrt{ \bar{m}_\chi^2 
- 4 \left(\bar{\xi}_{1,3} - \sqrt{\bar{\xi}_{1,3}\bar{\xi}_{2,3} } \right) \bar{m}_W^2 
- 4 \left(\bar{\xi}_{1,B} - \sqrt{\bar{\xi}_{1,B}\bar{\xi}_{2,B} } \right) \bar{m}_B^2 } 
\right) \, , \\
\label{mcWbarFull}
\bar{m}_{c_W}^2 &= \sqrt{\bar{\xi}_{1,W} \bar{\xi}_{2,W}} \bar{m}_W^2 \, , \\
\label{mcZhibarFull}
\bar{m}_{c_Z}^2 &= \sqrt{\bar{\xi}_{1,3} \bar{\xi}_{2,3}} \bar{m}_W^2 + \sqrt{\bar{\xi}_{1,B} \bar{\xi}_{2,B}} \bar{m}_B^2 \, . 
\end{align}
While for the gauge-dependent part of the one-loop anomalous dimension we get 
\begin{multline}
\label{anomaldimBCKGFull}
\left. \gamma^{(1)} \right|_{\rm{gauge \ dep.}}^{\rm{BKGD}}  = 
\frac{1}{(4 \pi)^2} \left( 
\frac{1}{2} \left( \bar{\xi}_{1,W} - 2 \sqrt{\bar{\xi}_{1,W} \bar{\xi}_{2,W}} \right) g^2
+ \frac{1}{4} \left( \bar{\xi}_{1,3} - 2 \sqrt{\bar{\xi}_{1,3} \bar{\xi}_{2,3}} \right) g^2 \right.  \\
\left. + \frac{1}{4} \left( \bar{\xi}_{1,B} - 2 \sqrt{\bar{\xi}_{1,B} \bar{\xi}_{2,B}} \right) g'^2
\right) \, .
\end{multline}
Notice that in the $\bar{\xi}_{1,\alpha} \rightarrow \bar{\xi}_{2,\alpha}$
limit ($\alpha= W,3,B$) and for $3 = W$ one reproduces the background $R_\xi$ gauge results  
in \eq{1loopEPBCKG} and \eq{RGE2lphiBCKD}, while for $\bar{\xi}_{2,\alpha} \rightarrow 0$ ($\alpha= W,3,B$) 
and $3 = W$ one obtains the expressions in \eq{1loopEPFermi} and \eq{RGE2lphi} for the Fermi gauge.

Let us finally point out that the SM effective potential in the standard $R_\xi$ gauge can be obtained by replacing 
\begin{equation}
\label{Rxirepl}
\bar{\xi}_{2,\alpha}^{1/2} \rightarrow \bar{\xi}_{2,\alpha}^{1/2} v/\phi \, , 
\end{equation} 
in the $\phi$-dependent mass terms of \eq{1loopEPbareBCKGFull},
where $\alpha=W,3,B$ and $v = \sqrt{m^2 / \lambda}$ denotes the tree-level vev of the Higgs doublet. 
In the limit relevant for the study of the SM vacuum stability, namely $\phi \gg v$, 
the $R_\xi$ gauge reduces to the Fermi gauge. 
On the other hand, the expression of the SM effective potential in the standard $R_\xi$ gauge 
is more suited for broken-phase calculations. 
  
\bibliographystyle{utphys.bst}
\bibliography{bibliography}

\providecommand{\href}[2]{#2}\begingroup\raggedright\begin{thebibliography}{10}

\bibitem{Aad:2012tfa}
{\bfseries ATLAS Collaboration} Collaboration, G.~Aad {\em et~al.},
  ``{Observation of a new particle in the search for the Standard Model Higgs
  boson with the ATLAS detector at the LHC},''
  \href{http://dx.doi.org/10.1016/j.physletb.2012.08.020}{{\em Phys.Lett.}
  {\bfseries B716} (2012) 1--29},
\href{http://arxiv.org/abs/1207.7214}{{\ttfamily arXiv:1207.7214 [hep-ex]}}.

\bibitem{Chatrchyan:2012ufa}
{\bfseries CMS Collaboration} Collaboration, S.~Chatrchyan {\em et~al.},
  ``{Observation of a new boson at a mass of 125 GeV with the CMS experiment at
  the LHC},'' \href{http://dx.doi.org/10.1016/j.physletb.2012.08.021}{{\em
  Phys.Lett.} {\bfseries B716} (2012) 30--61},
\href{http://arxiv.org/abs/1207.7235}{{\ttfamily arXiv:1207.7235 [hep-ex]}}.

\bibitem{Holthausen:2011aa}
M.~Holthausen, K.~S. Lim, and M.~Lindner, ``{Planck scale Boundary Conditions
  and the Higgs Mass},'' \href{http://dx.doi.org/10.1007/JHEP02(2012)037}{{\em
  JHEP} {\bfseries 1202} (2012) 037},
\href{http://arxiv.org/abs/1112.2415}{{\ttfamily arXiv:1112.2415 [hep-ph]}}.

\bibitem{EliasMiro:2011aa}
J.~Elias-Miro, J.~R. Espinosa, G.~F. Giudice, G.~Isidori, A.~Riotto, {\em
  et~al.}, ``{Higgs mass implications on the stability of the electroweak
  vacuum},'' \href{http://dx.doi.org/10.1016/j.physletb.2012.02.013}{{\em
  Phys.Lett.} {\bfseries B709} (2012) 222--228},
\href{http://arxiv.org/abs/1112.3022}{{\ttfamily arXiv:1112.3022 [hep-ph]}}.

\bibitem{Bezrukov:2012sa}
F.~Bezrukov, M.~Y. Kalmykov, B.~A. Kniehl, and M.~Shaposhnikov, ``{Higgs Boson
  Mass and New Physics},''
  \href{http://dx.doi.org/10.1007/JHEP10(2012)140}{{\em JHEP} {\bfseries 1210}
  (2012) 140},
\href{http://arxiv.org/abs/1205.2893}{{\ttfamily arXiv:1205.2893 [hep-ph]}}.

\bibitem{Degrassi:2012ry}
G.~Degrassi, S.~Di~Vita, J.~Elias-Miro, J.~R. Espinosa, G.~F. Giudice, {\em
  et~al.}, ``{Higgs mass and vacuum stability in the Standard Model at NNLO},''
  \href{http://dx.doi.org/10.1007/JHEP08(2012)098}{{\em JHEP} {\bfseries 1208}
  (2012) 098},
\href{http://arxiv.org/abs/1205.6497}{{\ttfamily arXiv:1205.6497 [hep-ph]}}.

\bibitem{Alekhin:2012py}
S.~Alekhin, A.~Djouadi, and S.~Moch, ``{The top quark and Higgs boson masses
  and the stability of the electroweak vacuum},''
  \href{http://dx.doi.org/10.1016/j.physletb.2012.08.024}{{\em Phys.Lett.}
  {\bfseries B716} (2012) 214--219},
\href{http://arxiv.org/abs/1207.0980}{{\ttfamily arXiv:1207.0980 [hep-ph]}}.

\bibitem{Masina:2012tz}
I.~Masina, ``{Higgs boson and top quark masses as tests of electroweak vacuum
  stability},'' \href{http://dx.doi.org/10.1103/PhysRevD.87.053001}{{\em
  Phys.Rev.} {\bfseries D87} no.~5, (2013) 053001},
\href{http://arxiv.org/abs/1209.0393}{{\ttfamily arXiv:1209.0393 [hep-ph]}}.

\bibitem{Buttazzo:2013uya}
D.~Buttazzo, G.~Degrassi, P.~P. Giardino, G.~F. Giudice, F.~Sala, {\em et~al.},
  ``{Investigating the near-criticality of the Higgs boson},''
  \href{http://dx.doi.org/10.1007/JHEP12(2013)089}{{\em JHEP} {\bfseries 1312}
  (2013) 089},
\href{http://arxiv.org/abs/1307.3536}{{\ttfamily arXiv:1307.3536}}.

\bibitem{Antipin:2013sga}
O.~Antipin, M.~Gillioz, J.~Krog, E.~M¿lgaard, and F.~Sannino, ``{Standard Model
  Vacuum Stability and Weyl Consistency Conditions},''
  \href{http://dx.doi.org/10.1007/JHEP08(2013)034}{{\em JHEP} {\bfseries 1308}
  (2013) 034},
\href{http://arxiv.org/abs/1306.3234}{{\ttfamily arXiv:1306.3234}}.

\bibitem{Branchina:2013jra}
V.~Branchina and E.~Messina, ``{Stability, Higgs Boson Mass and New Physics},''
  \href{http://dx.doi.org/10.1103/PhysRevLett.111.241801}{{\em Phys.Rev.Lett.}
  {\bfseries 111} (2013) 241801},
\href{http://arxiv.org/abs/1307.5193}{{\ttfamily arXiv:1307.5193 [hep-ph]}}.

\bibitem{Lindner:1988ww}
M.~Lindner, M.~Sher, and H.~W. Zaglauer, ``{Probing Vacuum Stability Bounds at
  the Fermilab Collider},''
\href{http://dx.doi.org/10.1016/0370-2693(89)90540-6}{{\em Phys.Lett.}
  {\bfseries B228} (1989) 139}.

\bibitem{Arnold:1989cb}
P.~B. Arnold, ``{Can the Electroweak Vacuum Be Unstable?},''
\href{http://dx.doi.org/10.1103/PhysRevD.40.613}{{\em Phys.Rev.} {\bfseries
  D40} (1989) 613}.

\bibitem{Sher:1988mj}
M.~Sher, ``{Electroweak Higgs Potentials and Vacuum Stability},''
\href{http://dx.doi.org/10.1016/0370-1573(89)90061-6}{{\em Phys.Rept.}
  {\bfseries 179} (1989) 273--418}.

\bibitem{Sher:1993mf}
M.~Sher, ``{Precise vacuum stability bound in the standard model},''
  \href{http://dx.doi.org/10.1016/0370-2693(93)91586-C}{{\em Phys.Lett.}
  {\bfseries B317} (1993) 159--163},
\href{http://arxiv.org/abs/hep-ph/9307342}{{\ttfamily arXiv:hep-ph/9307342
  [hep-ph]}}.

\bibitem{Ford:1992mv}
C.~Ford, D.~Jones, P.~Stephenson, and M.~Einhorn, ``{The Effective potential
  and the renormalization group},''
  \href{http://dx.doi.org/10.1016/0550-3213(93)90206-5}{{\em Nucl.Phys.}
  {\bfseries B395} (1993) 17--34},
\href{http://arxiv.org/abs/hep-lat/9210033}{{\ttfamily arXiv:hep-lat/9210033
  [hep-lat]}}.

\bibitem{Altarelli:1994rb}
G.~Altarelli and G.~Isidori, ``{Lower limit on the Higgs mass in the standard
  model: An Update},''
\href{http://dx.doi.org/10.1016/0370-2693(94)91458-3}{{\em Phys.Lett.}
  {\bfseries B337} (1994) 141--144}.

\bibitem{Casas:1994qy}
J.~Casas, J.~Espinosa, and M.~Quiros, ``{Improved Higgs mass stability bound in
  the standard model and implications for supersymmetry},''
  \href{http://dx.doi.org/10.1016/0370-2693(94)01404-Z}{{\em Phys.Lett.}
  {\bfseries B342} (1995) 171--179},
\href{http://arxiv.org/abs/hep-ph/9409458}{{\ttfamily arXiv:hep-ph/9409458
  [hep-ph]}}.

\bibitem{Espinosa:1995se}
J.~Espinosa and M.~Quiros, ``{Improved metastability bounds on the standard
  model Higgs mass},''
  \href{http://dx.doi.org/10.1016/0370-2693(95)00572-3}{{\em Phys.Lett.}
  {\bfseries B353} (1995) 257--266},
\href{http://arxiv.org/abs/hep-ph/9504241}{{\ttfamily arXiv:hep-ph/9504241
  [hep-ph]}}.

\bibitem{Casas:1996aq}
J.~Casas, J.~Espinosa, and M.~Quiros, ``{Standard model stability bounds for
  new physics within LHC reach},''
  \href{http://dx.doi.org/10.1016/0370-2693(96)00682-X}{{\em Phys.Lett.}
  {\bfseries B382} (1996) 374--382},
\href{http://arxiv.org/abs/hep-ph/9603227}{{\ttfamily arXiv:hep-ph/9603227
  [hep-ph]}}.

\bibitem{Isidori:2001bm}
G.~Isidori, G.~Ridolfi, and A.~Strumia, ``{On the metastability of the standard
  model vacuum},'' \href{http://dx.doi.org/10.1016/S0550-3213(01)00302-9}{{\em
  Nucl.Phys.} {\bfseries B609} (2001) 387--409},
\href{http://arxiv.org/abs/hep-ph/0104016}{{\ttfamily arXiv:hep-ph/0104016
  [hep-ph]}}.

\bibitem{Isidori:2007vm}
G.~Isidori, V.~S. Rychkov, A.~Strumia, and N.~Tetradis, ``{Gravitational
  corrections to standard model vacuum decay},''
  \href{http://dx.doi.org/10.1103/PhysRevD.77.025034}{{\em Phys.Rev.}
  {\bfseries D77} (2008) 025034},
\href{http://arxiv.org/abs/0712.0242}{{\ttfamily arXiv:0712.0242 [hep-ph]}}.

\bibitem{Ellis:2009tp}
J.~Ellis, J.~Espinosa, G.~Giudice, A.~Hoecker, and A.~Riotto, ``{The Probable
  Fate of the Standard Model},''
  \href{http://dx.doi.org/10.1016/j.physletb.2009.07.054}{{\em Phys.Lett.}
  {\bfseries B679} (2009) 369--375},
\href{http://arxiv.org/abs/0906.0954}{{\ttfamily arXiv:0906.0954 [hep-ph]}}.

\bibitem{Coleman:1973jx}
S.~R. Coleman and E.~J. Weinberg, ``{Radiative Corrections as the Origin of
  Spontaneous Symmetry Breaking},''
\href{http://dx.doi.org/10.1103/PhysRevD.7.1888}{{\em Phys.Rev.} {\bfseries D7}
  (1973) 1888--1910}.

\bibitem{Jackiw:1974cv}
R.~Jackiw, ``{Functional evaluation of the effective potential},''
\href{http://dx.doi.org/10.1103/PhysRevD.9.1686}{{\em Phys.Rev.} {\bfseries D9}
  (1974) 1686}.

\bibitem{Dolan:1974gu}
L.~Dolan and R.~Jackiw, ``{Gauge Invariant Signal for Gauge Symmetry
  Breaking},''
\href{http://dx.doi.org/10.1103/PhysRevD.9.2904}{{\em Phys.Rev.} {\bfseries D9}
  (1974) 2904}.

\bibitem{Kang:1974yj}
J.~Kang, ``{Gauge Invariance of the Scalar-Vector Mass Ratio in the
  Coleman-Weinberg Model},''
\href{http://dx.doi.org/10.1103/PhysRevD.10.3455}{{\em Phys.Rev.} {\bfseries
  D10} (1974) 3455}.

\bibitem{Fischler:1974ue}
W.~Fischler and R.~Brout, ``{Gauge Invariance in Spontaneously Broken
  Symmetry},''
\href{http://dx.doi.org/10.1103/PhysRevD.11.905}{{\em Phys.Rev.} {\bfseries
  D11} (1975) 905}.

\bibitem{Frere:1974ia}
J.-M. Frere and P.~Nicoletopoulos, ``{Gauge Invariant Content of the Effective
  Potential},''
\href{http://dx.doi.org/10.1103/PhysRevD.11.2332}{{\em Phys.Rev.} {\bfseries
  D11} (1975) 2332}.

\bibitem{Nielsen:1975fs}
N.~Nielsen, ``{On the Gauge Dependence of Spontaneous Symmetry Breaking in
  Gauge Theories},''
\href{http://dx.doi.org/10.1016/0550-3213(75)90301-6}{{\em Nucl.Phys.}
  {\bfseries B101} (1975) 173}.

\bibitem{Fukuda:1975di}
R.~Fukuda and T.~Kugo, ``{Gauge Invariance in the Effective Action and
  Potential},''
\href{http://dx.doi.org/10.1103/PhysRevD.13.3469}{{\em Phys.Rev.} {\bfseries
  D13} (1976) 3469}.

\bibitem{Aitchison:1983ns}
I.~Aitchison and C.~Fraser, ``{Gauge Invariance and the Effective Potential},''
\href{http://dx.doi.org/10.1016/0003-4916(84)90209-4}{{\em Annals Phys.}
  {\bfseries 156} (1984) 1}.

\bibitem{Johnston:1984sc}
D.~Johnston, ``{Nielsen Identities in the 't Hooft Gauge},''
\href{http://dx.doi.org/10.1016/0550-3213(85)90553-X}{{\em Nucl.Phys.}
  {\bfseries B253} (1985) 687}.

\bibitem{Thompson:1985hp}
G.~Thompson and H.-L. Yu, ``{Gauge covariance of the effective potential},''
\href{http://dx.doi.org/10.1103/PhysRevD.31.2141}{{\em Phys.Rev.} {\bfseries
  D31} (1985) 2141--2144}.

\bibitem{Kobes:1990dc}
R.~Kobes, G.~Kunstatter, and A.~Rebhan, ``{Gauge dependence identities and
  their application at finite temperature},''
\href{http://dx.doi.org/10.1016/0550-3213(91)90300-M}{{\em Nucl.Phys.}
  {\bfseries B355} (1991) 1--37}.

\bibitem{Ramaswamy:1995np}
S.~Ramaswamy, ``{Gauge invariance and the effective potential: The Abelian
  Higgs model},''
\href{http://dx.doi.org/10.1016/0550-3213(95)00331-L}{{\em Nucl.Phys.}
  {\bfseries B453} (1995) 240--258}.

\bibitem{Metaxas:1995ab}
D.~Metaxas and E.~J. Weinberg, ``{Gauge independence of the bubble nucleation
  rate in theories with radiative symmetry breaking},''
  \href{http://dx.doi.org/10.1103/PhysRevD.53.836}{{\em Phys.Rev.} {\bfseries
  D53} (1996) 836--843},
\href{http://arxiv.org/abs/hep-ph/9507381}{{\ttfamily arXiv:hep-ph/9507381
  [hep-ph]}}.

\bibitem{DelCima:1999gg}
O.~M. Del~Cima, D.~H. Franco, and O.~Piguet, ``{Gauge independence of the
  effective potential revisited},''
  \href{http://dx.doi.org/10.1016/S0550-3213(99)00226-6}{{\em Nucl.Phys.}
  {\bfseries B551} (1999) 813--825},
\href{http://arxiv.org/abs/hep-th/9902084}{{\ttfamily arXiv:hep-th/9902084
  [hep-th]}}.

\bibitem{Gambino:1999ai}
P.~Gambino and P.~A. Grassi, ``{The Nielsen identities of the SM and the
  definition of mass},''
  \href{http://dx.doi.org/10.1103/PhysRevD.62.076002}{{\em Phys.Rev.}
  {\bfseries D62} (2000) 076002},
\href{http://arxiv.org/abs/hep-ph/9907254}{{\ttfamily arXiv:hep-ph/9907254
  [hep-ph]}}.

\bibitem{Alexander:2008hd}
L.~P. Alexander and A.~Pilaftsis, ``{The One-Loop Effective Potential in
  Non-Linear Gauges},''
  \href{http://dx.doi.org/10.1088/0954-3899/36/4/045006}{{\em J.Phys.}
  {\bfseries G36} (2009) 045006},
\href{http://arxiv.org/abs/0809.1580}{{\ttfamily arXiv:0809.1580 [hep-ph]}}.

\bibitem{Loinaz:1997td}
W.~Loinaz and R.~Willey, ``{Gauge dependence of lower bounds on the Higgs mass
  derived from electroweak vacuum stability constraints},''
  \href{http://dx.doi.org/10.1103/PhysRevD.56.7416}{{\em Phys.Rev.} {\bfseries
  D56} (1997) 7416--7426},
\href{http://arxiv.org/abs/hep-ph/9702321}{{\ttfamily arXiv:hep-ph/9702321
  [hep-ph]}}.

\bibitem{Gonderinger:2012rd}
M.~Gonderinger, H.~Lim, and M.~J. Ramsey-Musolf, ``{Complex Scalar Singlet Dark
  Matter: Vacuum Stability and Phenomenology},''
  \href{http://dx.doi.org/10.1103/PhysRevD.86.043511}{{\em Phys.Rev.}
  {\bfseries D86} (2012) 043511},
\href{http://arxiv.org/abs/1202.1316}{{\ttfamily arXiv:1202.1316 [hep-ph]}}.

\bibitem{Ford:1992pn}
C.~Ford, I.~Jack, and D.~Jones, ``{The Standard model effective potential at
  two loops},'' \href{http://dx.doi.org/10.1016/0550-3213(92)90165-8}{{\em
  Nucl.Phys.} {\bfseries B387} (1992) 373--390},
\href{http://arxiv.org/abs/hep-ph/0111190}{{\ttfamily arXiv:hep-ph/0111190
  [hep-ph]}}.

\bibitem{Martin:2001vx}
S.~P. Martin, ``{Two loop effective potential for a general renormalizable
  theory and softly broken supersymmetry},''
  \href{http://dx.doi.org/10.1103/PhysRevD.65.116003}{{\em Phys.Rev.}
  {\bfseries D65} (2002) 116003},
\href{http://arxiv.org/abs/hep-ph/0111209}{{\ttfamily arXiv:hep-ph/0111209
  [hep-ph]}}.

\bibitem{Martin:2013gka}
S.~P. Martin, ``{Three-loop Standard Model effective potential at leading order
  in strong and top Yukawa couplings},''
  \href{http://dx.doi.org/10.1103/PhysRevD.89.013003}{{\em Phys.Rev.}
  {\bfseries D89} (2014) 013003},
\href{http://arxiv.org/abs/1310.7553}{{\ttfamily arXiv:1310.7553 [hep-ph]}}.

\bibitem{Patel:2011th}
H.~H. Patel and M.~J. Ramsey-Musolf, ``{Baryon Washout, Electroweak Phase
  Transition, and Perturbation Theory},''
  \href{http://dx.doi.org/10.1007/JHEP07(2011)029}{{\em JHEP} {\bfseries 1107}
  (2011) 029},
\href{http://arxiv.org/abs/1101.4665}{{\ttfamily arXiv:1101.4665 [hep-ph]}}.

\bibitem{Delaunay:2007wb}
C.~Delaunay, C.~Grojean, and J.~D. Wells, ``{Dynamics of Non-renormalizable
  Electroweak Symmetry Breaking},''
  \href{http://dx.doi.org/10.1088/1126-6708/2008/04/029}{{\em JHEP} {\bfseries
  0804} (2008) 029},
\href{http://arxiv.org/abs/0711.2511}{{\ttfamily arXiv:0711.2511 [hep-ph]}}.

\bibitem{Collins:1984xc}
J.~C. Collins, ``{Renormalization. An Introduction to Renormalization, the
  Renormalization Group, and the Operator Product Expansion},''
{\em Cambridge Monographs on Mathematical Physics} (1984).

\bibitem{Bardeen:1978yd}
W.~A. Bardeen, A.~Buras, D.~Duke, and T.~Muta, ``{Deep Inelastic Scattering
  Beyond the Leading Order in Asymptotically Free Gauge Theories},''
\href{http://dx.doi.org/10.1103/PhysRevD.18.3998}{{\em Phys.Rev.} {\bfseries
  D18} (1978) 3998}.

\bibitem{Chetyrkin:2012rz}
K.~Chetyrkin and M.~Zoller, ``{Three-loop $\beta$-functions for top-Yukawa and
  the Higgs self-interaction in the Standard Model},''
  \href{http://dx.doi.org/10.1007/JHEP06(2012)033}{{\em JHEP} {\bfseries 1206}
  (2012) 033},
\href{http://arxiv.org/abs/1205.2892}{{\ttfamily arXiv:1205.2892 [hep-ph]}}.

\bibitem{Mihaila:2012pz}
L.~N. Mihaila, J.~Salomon, and M.~Steinhauser, ``{Renormalization constants and
  beta functions for the gauge couplings of the Standard Model to three-loop
  order},'' \href{http://dx.doi.org/10.1103/PhysRevD.86.096008}{{\em Phys.Rev.}
  {\bfseries D86} (2012) 096008},
\href{http://arxiv.org/abs/1208.3357}{{\ttfamily arXiv:1208.3357 [hep-ph]}}.

\bibitem{Pilaftsis:1997fe}
A.~Pilaftsis, ``{Higgs boson low-energy theorem and compatible gauge fixing
  conditions},'' \href{http://dx.doi.org/10.1016/S0370-2693(97)01516-5}{{\em
  Phys.Lett.} {\bfseries B422} (1998) 201--211},
\href{http://arxiv.org/abs/hep-ph/9711420}{{\ttfamily arXiv:hep-ph/9711420
  [hep-ph]}}.

\bibitem{Binosi:2005yk}
D.~Binosi, J.~Papavassiliou, and A.~Pilaftsis, ``{Displacement operator
  formalism for renormalization and gauge dependence to all orders},''
  \href{http://dx.doi.org/10.1103/PhysRevD.71.085007}{{\em Phys.Rev.}
  {\bfseries D71} (2005) 085007},
\href{http://arxiv.org/abs/hep-ph/0501259}{{\ttfamily arXiv:hep-ph/0501259
  [hep-ph]}}.

\bibitem{Appelquist:1973ms}
T.~Appelquist, J.~Carazzone, J.~T. Goldman, and H.~R. Quinn, ``{Renormalization
  and gauge independence in spontaneously broken gauge theories},''
\href{http://dx.doi.org/10.1103/PhysRevD.8.1747}{{\em Phys.Rev.} {\bfseries D8}
  (1973) 1747--1756}.

\bibitem{Sirlin:1985ux}
A.~Sirlin and R.~Zucchini, ``{Dependence of the Quartic Coupling H(m) on M($H$)
  and the Possible Onset of New Physics in the Higgs Sector of the Standard
  Model},''
\href{http://dx.doi.org/10.1016/0550-3213(86)90096-9}{{\em Nucl.Phys.}
  {\bfseries B266} (1986) 389}.

\bibitem{Weinberg:1987vp}
E.~J. Weinberg and A.-q. Wu, ``{Understanding Complex Perturbative Effective
  Potentials},''
\href{http://dx.doi.org/10.1103/PhysRevD.36.2474}{{\em Phys.Rev.} {\bfseries
  D36} (1987) 2474}.

\bibitem{Einhorn:1980ik}
M.~B. Einhorn and K.~Sato, ``{Monopole Production in the Very Early Universe in
  a First Order Phase Transition},''
\href{http://dx.doi.org/10.1016/0550-3213(81)90057-2}{{\em Nucl.Phys.}
  {\bfseries B180} (1981) 385}.

\bibitem{Hung:1995in}
P.~Hung and M.~Sher, ``{Implications of a Higgs discovery at LEP},''
  \href{http://dx.doi.org/10.1016/0370-2693(96)00123-2}{{\em Phys.Lett.}
  {\bfseries B374} (1996) 138--144},
\href{http://arxiv.org/abs/hep-ph/9512313}{{\ttfamily arXiv:hep-ph/9512313
  [hep-ph]}}.

\bibitem{Casas:2000mn}
J.~Casas, V.~Di~Clemente, and M.~Quiros, ``{The Standard model instability and
  the scale of new physics},''
  \href{http://dx.doi.org/10.1016/S0550-3213(00)00199-1}{{\em Nucl.Phys.}
  {\bfseries B581} (2000) 61--72},
\href{http://arxiv.org/abs/hep-ph/0002205}{{\ttfamily arXiv:hep-ph/0002205
  [hep-ph]}}.

\bibitem{Ade:2014xna}
{\bfseries BICEP2} Collaboration, P.~Ade {\em et~al.}, ``{BICEP2 I: Detection
  Of B-mode Polarization at Degree Angular Scales},''
\href{http://arxiv.org/abs/1403.3985}{{\ttfamily arXiv:1403.3985
  [astro-ph.CO]}}.

\bibitem{Espinosa:2007qp}
J.~Espinosa, G.~Giudice, and A.~Riotto, ``{Cosmological implications of the
  Higgs mass measurement},''
  \href{http://dx.doi.org/10.1088/1475-7516/2008/05/002}{{\em JCAP} {\bfseries
  0805} (2008) 002},
\href{http://arxiv.org/abs/0710.2484}{{\ttfamily arXiv:0710.2484 [hep-ph]}}.

\bibitem{Kobakhidze:2013tn}
A.~Kobakhidze and A.~Spencer-Smith, ``{Electroweak Vacuum (In)Stability in an
  Inflationary Universe},''
  \href{http://dx.doi.org/10.1016/j.physletb.2013.04.013}{{\em Phys.Lett.}
  {\bfseries B722} (2013) 130--134},
\href{http://arxiv.org/abs/1301.2846}{{\ttfamily arXiv:1301.2846 [hep-ph]}}.

\bibitem{Fairbairn:2014zia}
M.~Fairbairn and R.~Hogan, ``{Electroweak Vacuum Stability in light of
  BICEP-2},''
\href{http://arxiv.org/abs/1403.6786}{{\ttfamily arXiv:1403.6786 [hep-ph]}}.

\bibitem{Enqvist:2014bua}
K.~Enqvist, T.~Meriniemi, and S.~Nurmi, ``{Higgs Dynamics during Inflation},''
\href{http://arxiv.org/abs/1404.3699}{{\ttfamily arXiv:1404.3699 [hep-ph]}}.

\bibitem{Kobakhidze:2014xda}
A.~Kobakhidze and A.~Spencer-Smith, ``{The Higgs vacuum is unstable},''
\href{http://arxiv.org/abs/1404.4709}{{\ttfamily arXiv:1404.4709 [hep-ph]}}.

\bibitem{Hook:2014uia}
A.~Hook, J.~Kearney, B.~Shakya, and K.~M. Zurek, ``{Probable or Improbable
  Universe? Correlating Electroweak Vacuum Instability with the Scale of
  Inflation},''
\href{http://arxiv.org/abs/1404.5953}{{\ttfamily arXiv:1404.5953 [hep-ph]}}.

\bibitem{Bednyakov:2012rb}
A.~Bednyakov, A.~Pikelner, and V.~Velizhanin, ``{Anomalous dimensions of gauge
  fields and gauge coupling beta-functions in the Standard Model at three
  loops},'' \href{http://dx.doi.org/10.1007/JHEP01(2013)017}{{\em JHEP}
  {\bfseries 1301} (2013) 017},
\href{http://arxiv.org/abs/1210.6873}{{\ttfamily arXiv:1210.6873 [hep-ph]}}.

\bibitem{Chetyrkin:2013wya}
K.~Chetyrkin and M.~Zoller, ``{$\beta$-function for the Higgs self-interaction
  in the Standard Model at three-loop level},''
  \href{http://dx.doi.org/10.1007/JHEP04(2013)091,
  10.1007/JHEP09(2013)155}{{\em JHEP} {\bfseries 1304} (2013) 091},
\href{http://arxiv.org/abs/1303.2890}{{\ttfamily arXiv:1303.2890 [hep-ph]}}.

\bibitem{Bednyakov:2013eba}
A.~Bednyakov, A.~Pikelner, and V.~Velizhanin, ``{Higgs self-coupling
  beta-function in the Standard Model at three loops},''
  \href{http://dx.doi.org/10.1016/j.nuclphysb.2013.07.015}{{\em Nucl.Phys.}
  {\bfseries B875} (2013) 552--565},
\href{http://arxiv.org/abs/1303.4364}{{\ttfamily arXiv:1303.4364}}.

\bibitem{Bohm:1986rj}
M.~Bohm, H.~Spiesberger, and W.~Hollik, ``{On the One Loop Renormalization of
  the Electroweak Standard Model and Its Application to Leptonic Processes},''
\href{http://dx.doi.org/10.1002/prop.19860341102}{{\em Fortsch.Phys.}
  {\bfseries 34} (1986) 687--751}.

\end{thebibliography}\endgroup

\end{document}